\documentclass[a4paper,11pt]{article}
\usepackage{amsmath,amssymb,mathtools}
\usepackage{color}
\usepackage{graphicx}
\usepackage{subfigure}
\usepackage{tikz}
\usepackage{xparse}
\usepackage{appendix}
\usepackage{hyperref}
\usepackage{mathrsfs}
\usepackage{amsthm}
\usepackage{float} 
\usepackage{ulem}
\usepackage[text={17cm,24.5cm},centering]{geometry}
\def \be {\begin{equation}}
\def \ee {\end{equation}}
\def \ba {\begin{array}}
\def \ea {\end{array}}
\def \bea{\begin{eqnarray}}
\def \eea{\end{eqnarray}}
\def \nn {\nonumber}

\def \Trho {\langle T \rangle_\rho}
\def \Arho {\langle \mathcal A \rangle_\rho}
\def \Brho {\langle \mathcal B \rangle_\rho}
\def \Drho {\langle \mathcal D \rangle_\rho}
\def \mA {\mathcal A}
\def \mB {\mathcal B}

\def \mD {\mathcal D}

\begin{document}

\title{Correlations in geometric states}
\author{Wu-zhong Guo\footnote{wuzhong@hust.edu.cn}~}

\date{}
\maketitle

\vspace{-10mm}
\begin{center}
{\it School of Physics, Huazhong University of Science and Technology,\\
 Wuhan, Hubei
430074, China
\vspace{1mm}
}
\vspace{10mm}
\end{center}

\begin{abstract}
In this paper we explore the correlations in the geometric states. Here the geometric state means the state in CFTs that can be effectively described by classical geometry in the bulk in the semi-classical limit $G\to 0$. By using  the upper bound of Holevo informaion we show the covex combination of geometric states cannot be a geometric state. To understand the duality between thermofield double state and eternal black hle, we construct several correlated states of two CFTs. In all the examples we show their correlations are too weak to produce the a connected spacetime. we review the measure named quantum discord and use it to characterize the classical and quantum correlations in quantum field theories.  Finally,  we discuss the correlations between two intervals $A$ and $B$ with distance $d$ in the vacuum state of 2D CFTs with large central charge $c$.  The feature is the phase transition of the mutual information $I(\rho_{AB})$.  We analyse the quasi-product state of $\rho_{AB}$ for large $d$. By using the Koashi-Winter relation of tripartite states the quantum and classical correlations between $A$ and $B$ can expressed as Holevo information, which provides a new understanding of the correlations as accessible information. 
\end{abstract}

\newpage

\section{Introduction}
For a given quantum state, generally the structures of multipartite correlations are complicated, which include classical correlation and quantum correlation. Entanglement is the main feature of the quantum correlation but not the only.  Usually,  it is not an easy task to characterize and quantify the quantum correlation even for bipartite states, see, e.g., the  review  \cite{Horodecki:2009zz}.\\
For quantum field theories (QFTs) the correlations between different regions are closely related to the intrinsic parameters of the theories.  For particle physics the correlation functions are used to detect the parameters of the underlying theory. The theories with gravity dual are special since it is found some measures of entanglement may be associated with some geometric quantities. The dual relations open us  a new way to explore the structure of correlations in QFTs as well as emergence of spacetime of gravity theories.  \\
In the context of AdS/CFT the celabrated  Ryu-Takayanagi (RT) formula \cite{Ryu:2006bv}  show the entanglement entropy (EE) of a subsystem $A$ is associated with a minimal surface in the bulk homologous to the $A$.  
Based on RT formula many fruitful results have given us more deep understanding on AdS/CFT and spacetime emergence, see a recent review on this direction\cite{Rangamani:2016dms}. 
The entanglement wedge defined as the region surrounded by the subsystem $A$ and its minimal surface  provides a nature understanding of subregion/subregion duality\cite{Czech:2012bh}-\cite{Dong:2016eik}. For two subsystems $A$ and $B$  the minimal cross of entanglement wedge is an interesting geometric quantity in the bulk.  It is conjectured to be associated with entanglement of purification (EoP) of state $\rho_{AB}$ \cite{Takayanagi:2017knl}\cite{Nguyen:2017yqw}, see also other possible conjectures\cite{Kudler-Flam:2018qjo}-\cite{Dutta:2019gen}. \\

A notable fact is that the proposal of the gravity dual of some special measures should be only right for the geometric states. We define the geometric states as the states in conformal field thoeries (CFTs) that can be described by a classical geometry in the semi-classical limit $G\to 0$, or the large central charge limit $c\to \infty$. So an important question is how to judge whether a state is a geometric one or not. At present there is no short criterion on this problem\cite{Rangamani:2016dms}. \\

 In fact we only know very few examples that are gemetric or not. In the paper one of our motivation is to study the set of geometric states. Specially, we focus on the problem that whether the set of geometric states is convex. The short answer is no. Our method is using the Holevo information which is used as accessible information of a given ensemble. It can also be a measure of the distinguishability among the microstates of the ensemble.  We estimate the upper bound of the a convex combination of geometric states denoted by $\rho_c^g$, and find it is at most be $\frac{1}{\sqrt{G}}$ or $\sqrt{c}$.  This means the holographic entanglement entropy is quasi-linear.  But if the state $\rho_c^g$ is a geometric state, we may use the RT formula to calculate the EE, the result is that the Holevo information is not vanishing. Therefore, we expect the assumption that $\rho_c^g$ is a geometric state is not right.\\

By the argument from \cite{VanRaamsdonk:2010pw} one expect the entanglement is important for the emergence of connected spacetime.  Entanglement is the main feature of quantum correlation, but it is not the only source of quantum correlation. Even the separable state, it is possible to have quantum correlation measured by quantum discord\cite{OZ}\cite{LV}. To catch the feature of correlations in geometric states, it is useful to discuss classical and quantum correlation respectively rather than just talking about entanglement.  Using the eternal black hole as an example, we construct states with less correlations than the thermofield double state.  One of the example is a state with only classical correlation. We find the correlator with operators inserted at the opposite boundary of eternal black hole is vanishing in the thermodynamic limit $L\to \infty$, where $L$ is the  spatial size of the boundary. The other examples have quantum correlations, but the correlator is also vanishing.  This suggests these states have less quantum correlations  than the thermofield double state thus fail to be dual to the eternal black hole. \\

In QFTs the quantum correlation between arbitrary regions is very general as a consequence of  the Reeh-Schlieder property of the vaccum.  Consider two subregions $A$ and $B$ with spacelike distance $d$. For large enough d, we would expect the two subregions will lose correlation, i.e., $\rho_{AB}= \rho_A\otimes \rho_B+\delta\rho_{AB}$, where $\delta\rho_{AB}$ is a small perturbation depending on the distance $d$.  The feature of the CFTs with gravity dual is that there exists a critical point of $d$  such that the correlation has a phase transition.  We disccuss this phenomenon in the vacuum  for 2D CFTs with large central charge $c$.   The mutual informaiton of $\rho_{AB}$ seems to be sensitive to the perturbation of the spectrum near $e^{-2(b_{A}+b_B)}$ where $b_{A(B)}=\frac{c}{6}\log l_{A(B)}/\epsilon$. \\

Further on $A$ and $B$, we derive some relations between the correlations between $A$,$B$ and its complementary. These relations involve of the measures quantum discord and entanglement of formation. Finally the quanutm correlation and classical correlation between $A$ and $B$ are expressed as the Holevo information, which is taken as the accessible information of a given ensemble. \\

This paper is organized as follows. We firstly discuss the set of geometric state. In the section.\ref{convexgeometry} by using the upper bound of Holevo information we derive the holographic entanglement entropy should be quasi-linear. But once assumed the convex combination of geometric state is a new geometric state, one would get an inconsistent result. This suggests the set of geometric state is not convex.  In section.\ref{Holevothermal} we review the analytical results of Holevo information of an interval with length $\ell$ in 2D CFT with the canonical and microcanonical ensemble thermal state.  We further show the Holevo information is vanishing in the large $L$ limit up to order $O(\ell^{12})$.  The results in this section will be used in next section. Then we explore the correlations in geometric states. For the example of eternal black hole, we construct several states which has less correlation than thermofield double state. These examples show the large quantum correlation is important for a geometric state. In section.\ref{classification} we review the operational measure to quantify quantum and classical correlation and classification of states by correlation. We also point out the existence of quantum correlation is general in QFTs states satisfying Reeh-Schlieder property.  In the section.\ref{final} we mainly focus on the example of two intervals in 2D CFTs.   An interesting feature is the phase transition of the mutual information depending on the distance between this two intervals. We find some special property of the mutual information in large central charge limit.  The mutual information between the two intervals can be expressed as the Holevo information. Conlcusion is in section.\ref{conclusion}.

\section{Geometric states}\label{geometricsection}
We call a state in CFT to be geometric if it can be effectively described by a classical  geometry, or more precise, the expectation values of observables or equally correlators in this state are consistent with the holographic results in the semiclassical limit $G\to 0$.  Otherwise we call the states to be non-geometric.   In \cite{Guo:2018fnv} the authors have shown some non-geometric states in  2D CFT with large central charge $c$, such as some special descendant states or the superposition of two geometric pure states. But at present we have no effective method to test whether a given state is geometric or not. \\
Given a set of states  $\rho_i$ with $i=1,2,...,n$,  one can construct a new density $\rho$ matrix by convex combination, 
\be\label{ConvexState}
\rho:= \sum_i p_i \rho_i,
\ee
with $0\le p_i\le 1$ and $\sum_i p_i =1$. For the state (\ref{ConvexState}) we can define  an useful upper bound of the accessible information of an observer who can perform any measurement on this state, named Holevo bound or Holevo information.  In this section we will use Holevo information to get some interesting properties for geometric states.
\subsection{Holevo Information and its upper bound}
The Holevo information denoted by $\chi$ is defined as
\be\label{holevodefinition}
\chi(\rho):= S(\rho)-\sum p_i S(\rho_i),
\ee
where $S(\rho)$ is the Von Neumann entropy of the state $\rho$.  By using the relative entropy 
\be
S(\sigma ||\rho):=tr (\sigma \log \sigma)-tr (\sigma \log \rho)
\ee
for two states $\sigma$ and $\rho$,
one may rewrite the Holevo information as the averge relative enropy, 
\bea
\chi(\rho)=\sum_i p_i S(\rho_i|| \rho).
\eea
Since the relative entropy is one kind of distance measure between two states,  the Holevo information can also be used to characterize the distinguishability of the states $\rho_i$. Specially,  if $\rho_i = | i \rangle \langle i |$  with $\langle i |j\rangle =\delta_{ij}$, we can perfectly distinguish the states $\rho_i$, in this case the Holevo information  $\chi(\rho)=S(\rho)$.  While for $\rho_i=\rho$ one cannot distinguish the states $\rho_i$ by any measurement,  in this case we have $\chi(\rho)=0$.\\
The Holevo information $\chi(\rho)$ satisfies the following inequality 
\bea\label{1bound}
0\le \chi(\rho) \le H(p_i),
\eea
where $H(p_i):= -\sum_i p_i \log p_i$ is the Shannon entropy of the probability $p_i$. A lower upper bound of $\chi(\rho)$ is derived in \cite{AKM}.  Let $T(\rho_i,\rho_j):=\frac{1}{2} ||\rho_i-\rho_j||_1$, where $T(\rho_i,\rho_j)$ is the trace distance between the two states $\rho_i$ and $\rho_j$. The factor $\frac{1}{2}$ ensures $0\le T(\rho_i,\rho_j)\le 1$.  Let $t=\text{max}_{\rho_i,\rho_j}T(\rho_i,\rho_j)$, we have the following bound 
\bea\label{bound}
\chi(\rho)\le H(p_i ) t.
\eea
The new upper bound is much lower than (\ref{1bound}) for the case $t$ is very small.
\subsection{Two qubits example}
\begin{figure}[H]
\centering 
\includegraphics[trim = 0mm 0mm 0mm 0mm, clip=true,width=10.0cm]{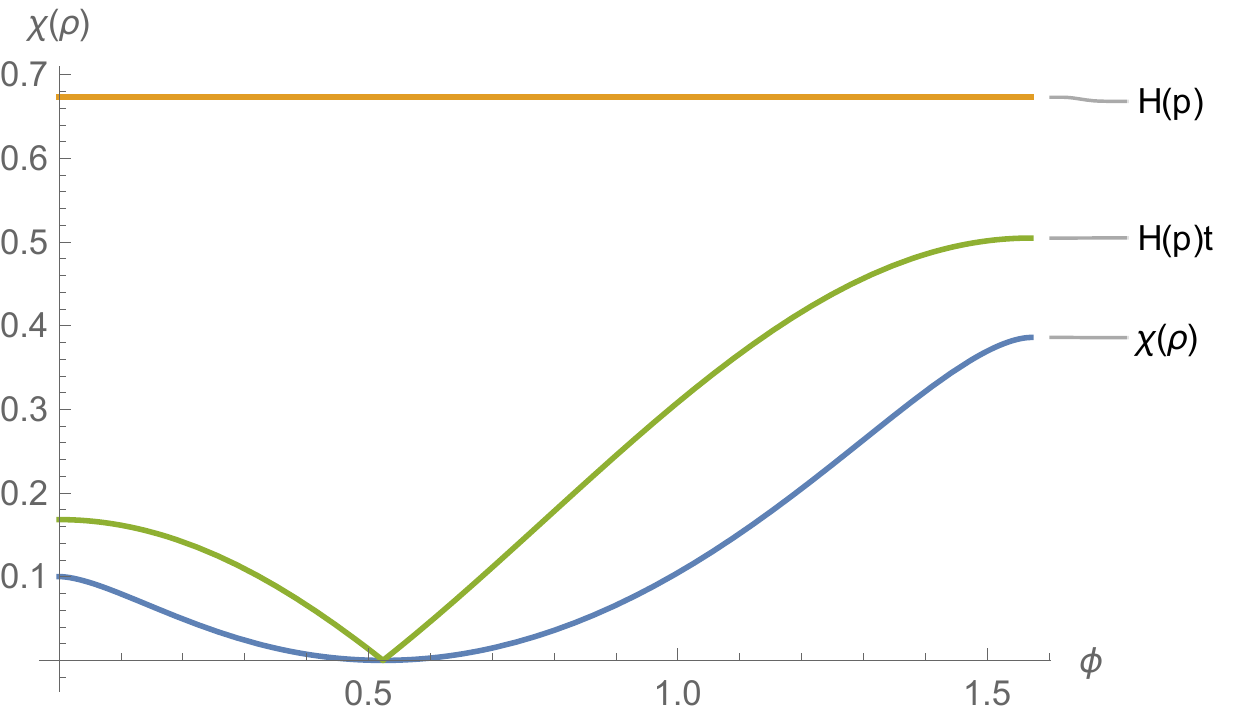}
\caption{Plot for the Holevo information and its two upper bound.}
\label{Holevoplot}
\end{figure}

For two qubits system, we have the basis $|0\rangle$ and $|1\rangle$. Let's consider the following state
\bea
\rho=p \rho_1+(1-p)\rho_2.
\eea
with $0\le p \le 1$ and
\bea
\rho_1=\cos^2(\theta)|0\rangle\langle 0|+\sin^2(\theta)|1\rangle\langle1\rangle,\quad \rho_2=\cos^2(\phi)|0\rangle\langle 0|+\sin^2(\phi)|1\rangle\langle1\rangle,
\eea
where $0\le \theta ,\phi\le 2\pi$.  By definition we get the Holevo information 
\bea
&&\chi(\rho)=-\left[p \cos ^2(\theta )+(1-p) \cos ^2(\phi )\right] \log \left[p \cos ^2(\theta )+(1-p) \cos ^2(\phi )\right]\nonumber \\
&&\phantom{\chi(\rho)=}-\left[p \sin ^2(\theta )+(1-p) \sin ^2(\phi )\right] \log \left[p \sin ^2(\theta )+(1-p) \sin ^2(\phi )\right]\nn \\
&&\phantom{\chi(\rho)=}+2 p \left(\sin ^2(\theta ) \log [\sin (\theta )]+\cos ^2(\theta ) \log [\cos (\theta )]\right)\nn \\
&&\phantom{\chi(\rho)=}+2 (1-p) \left(\sin ^2(\phi ) \log [\sin (\phi )]+\cos ^2(\phi ) \log [\cos (\phi )]\right).
\eea
The trace distance between $\rho_1$ and $\rho_2$ is
\bea
t=T(\rho_1,\rho_2)=\frac{1}{2}\sqrt{(\rho_1-\rho_2)^2}=|\cos^2(\theta)-\cos^2(\phi)|.
\eea
If the distance between $\rho_1$ and $\rho_2$ is small, e.g., taking $\phi=\theta+\epsilon$ with $\epsilon\ll 1$, we have 
\bea
\chi(\rho)=2p(1-p)\epsilon^2+O(\epsilon^3), \quad t=2|\cos(\theta)\sin(\theta)| \epsilon+O(\epsilon^2).
\eea
In this case $H(p)=-p\log(p)-(1-p)\log(1-p)$ is a very bad upper bound for $\chi(\rho)$, but $H(p)t$, which is $O(\epsilon)$, gives a much better bound for the Holevo information. We plot the Holevo information and its upper bound in Fig.\ref{Holevoplot}.
 \subsection{The set of geometric states}\label{convexgeometry}
In this section we will use the Holevo information to study the set of geometric states.  
Support the set of the geometric states $\mathcal{G}=\{\rho^g_i\}$ ($i=1,2,...$), we would like to show the convex combination of the geometric states cannot be a geometric state, that is
\bea\label{rhocon}
\rho_c=\sum_i p_i\rho^g_i \notin \mathcal{G},
\eea
where $\sum_i p_i =1$.
\subsubsection{General dimension}\label{higherdimension}
The general asymptotic $d+1$ dimensional bulk solutions can be formulated in the Fefferman-Graham (FG) coordinate, the metric is 
\bea\label{FG}
ds^2=\frac{L^2}{z^2}\left( dz^2+g_{\mu\nu }(z,x^\mu)dx^\mu dx^\nu \right),
\eea
where $g_{\mu\nu}=\eta_{\mu\nu}+z^d t_{\mu\nu}+O(z^{d+1})$. Here we only consider the CFT lives on the flat spacetime at the boundary $z=0$. The stress-energy tensor $T_{\mu\nu}$ of the dual state $\rho$ is associated with $t_{\mu\nu}$  with
\bea
\langle T_{\mu\nu}\rangle_\rho=\frac{d L^{d-1}}{16\pi G}t_{\mu\nu}.
\eea
The stress energy tensor in the state $\rho_c$ is
\bea
\langle T_{\mu\nu}\rangle_{\rho_c}=\sum_i p_i \langle T_{\mu\nu}\rangle_{\rho^g_i}=\frac{d L^{d-1}}{16\pi G}\sum_i p_i t_{i,\mu\nu},
\eea
where $ t_{i,\mu\nu}$ is the metric perturbation dual to the geometric state $\rho_i^g$. \\
Now let's consider the entanglement entropy for subsystem on the time slice $t=t_0$. For our purpose taking the subsystem $A$ to be a sphere with radius $R$.  The tensor $t_{\mu\nu}$ also has an energy density scale $m$ which is associated with energy scale of the perturbation. We can always choose  the size $R$ is very small, i.e., $R^{d}\ll 1/m$. Since the size of the sphere can be arbitrary small,  in this subregion we can take the stress-energy tensor to be constant.  In this limit we can perturbatively calculate the entanglement entropy of $A$ .  In general, the entanglement entropy of $A$ in the geometric state $\rho$ would have the following form, 
\bea\label{HEEperturbation1}
S(\rho_A)=S(\rho_{0,A})+ R^d S_1 +R^{2d} S_2+O(R^{3d}),
\eea
where  $\rho_{0,A}:= tr_{\bar A} |0\rangle \langle 0|$\footnote{In this expansion we only write explicitly  the parameter $R$, in fact $S_1$ and $S_2$ contain the scale $m$. This expansion is available only if the dimensionless parameter $R^d m\ll 1$.}.  The first order correction $S_1$ is proportional to the energy density $\langle T_{00}\rangle_\rho$, while the second order correction $S_2$  is related to $\langle T_{00}\rangle_\rho^2, \langle T_{0k}\rangle_\rho^2, \langle T_{kl}\rangle_\rho^2$. More precisely, we have
\bea\label{HEEperturbation2}
S_1=\frac{c_1}{16\pi G } \langle T_{00}\rangle_\rho, \quad S_2=-\frac{1}{16 \pi G}\big(c_2 \langle T_{00}\rangle_\rho^2+c_3 \langle T_{0k}\rangle_\rho^2 +c_4 \langle T_{kl}\rangle_\rho^2\big),
\eea
in which $c_1,c_2,c_3,c_4\ge 0$. One can get the above conclusion by using relative entropy $S(\rho_A|||\rho_{0,A})$. We can write the relative entropy as
\bea
S(\rho_A|||\rho_{0,A})=\Delta \langle H\rangle-\Delta S,
\eea
with
\bea
\Delta S= S(\rho_A)-S(\rho_{0,A}),\quad \Delta \langle H\rangle=\langle H \rangle_\rho-\langle H\rangle_{\rho_0},
\eea
where $H=2\pi \int_{A} d^{d-1}x \frac{R^2-|\bf{x}|^2}{2R}T_{00}$ is the modular Hamiltonian of the sphere $A$. To guarantee the positivity of relative entropy, the term of  order $O(R)$  should be vanishing and $S_2\ge 0$.  Using these  we fix the sign of the constants $c_m$ ($m=1,2,3,4$). \\
We also have 
\bea\label{relativeentropysecond}
S(\rho_A|||\rho_{0,A})=\frac{R^{2d}}{16\pi G}C_\rho+O(\frac{R^{3d}}{16\pi G}), \quad \text{with} \quad C_\rho=c_2 \langle T_{00}\rangle_\rho^2+c_3 \langle T_{0k}\rangle_\rho^2 +c_4 \langle T_{kl}\rangle_\rho^2.
\eea
In \cite{Blanco:2013joa} the authors have calculated the second order correction $S_2$ by using the RT formula for a constant stress-energy tensor. The results show $c_m$ ($m=1,2,3,4$) are constants independent with the state $\rho$.\\
The reduced density matrix $\rho_{c,A}$ of subsystem $A$ in the state $\rho_c$ (\ref{rhocon}) is
\bea
\rho_{c,A}=\sum_i p_i \rho^g_{i,A},
\eea
where $\rho^g_{i,A}=tr_{\bar A} \rho_i^g$.
The Holevo information of the above state is 
\bea
\chi(\rho_{c,A})=S(\sum_i p_i \rho^g_{i,A})-\sum_i p_i S( \rho^g_{i,A}).
\eea
By using the bound of Holevo information (\ref{1bound}) (\ref{bound}),
\bea
\chi(\rho_{c,A})\le \{ H(p_i), H(p_i)t\},
\eea
where $t$ is the maximal trace distance of the states in $\mathcal{G}$.  In general $H(p_i)t$ will give a better bound for the Holevo information if the distance among the states$\{ \rho^g_{i,A}\}$ is small. Specially here we consider the  entropy of a small subsystem $A$ the Holevo information is at least of order $O(R^dm)$, which is much smaller than $H(p_i)$. So we will focus on the bound $H(p_i)t $ with
\bea
t=T(\rho^g_{i_1,A},\rho^g_{i_2,A}).
\eea
The calculation of the trace distance in QFT is usually difficult, see some results in 2D CFTs \cite{Zhang:2019wqo}.  Here we can use the Pinsker's inequality to estimate the trance distance between the reduced density matrix of two geometric states.  By using the triangle inequality and the Pinsker's inequality we have
\bea
T(\rho^g_{i_1,A},\rho^g_{i_2,A})\le T(\rho^g_{i_1,A},\rho_{0,A})+T(\rho^g_{i_2,A},\rho_{0,A})\le \frac{1}{\sqrt{2}}(\sqrt{S(\rho^g_{i_1,A},,\rho_{0,A})}+\sqrt{S(\rho^g_{i_2,A},,\rho_{0,A})}).
\eea
Further using the result (\ref{relativeentropysecond}) we find
\bea
\chi_{\rho_{c,A}}\lesssim \frac{1}{\sqrt{32\pi G}}\big( \alpha_{i_1} \sqrt{C_{{\rho^g_{i_1}}}}+\alpha_{i_2} \sqrt{C_{{\rho^g_{i_2}}}}\big) +O(\alpha^2_{i_1},\alpha^2_{i_2}),
\eea
where $\alpha_{i_1},\alpha_{i_1}$ and $C_{{\rho^g_{i_1}}},C_{{\rho^g_{i_2}}}$ is defined by replacing the state $\rho$ by the respective states $\rho^g_{i_1}$ and $\rho^g_{i_2}$.  This means the Holevo information is at most $O(\frac{1}{\sqrt{G}})$, or equally,
\bea\label{linearcondition}
S(\rho_{c,A})= \sum_i p_i S( \rho^g_{i,A})+O(\frac{1}{\sqrt{G}}).
\eea
Since $S( \rho^g_{i,A})$ are all of $O(\frac{1}{G})$, in the semiclassical limit $G\to 0$ we may ignore the term at the order $O(\frac{1}{\sqrt{G}})$. This means the holographic entanglement entropy is quasi-linear.\\
Let's come back to the FG coordinate (\ref{FG}), which  is the most general solution of Einstein equation. The constraint on the tensor $t_{\mu\nu}$ by the Einstein equation is 
\bea\label{stressenergycondition}
t_\mu^\mu =0,\quad  \nabla^\mu t_{\mu\nu}=0,
\eea
which are associated with the  traceless and conserve condition of stress-energy tensor for CFTs. The tensor $t_{\mu\nu}$ corresponding to the state  $\rho_c$ also satisfies the conditions (\ref{stressenergycondition}). It seems the state $\rho_c$ is also a geometric state, i.e., it can be described by the FG coordinate with 
\bea
t_{\mu\nu} =\frac{16\pi G}{d L^{d-1}} \langle T_{\mu\nu}\rangle_{\rho_c}.
\eea
By using (\ref{HEEperturbation1}) (\ref{HEEperturbation2})  we find 
\bea
S(\rho_{c,A})\ne \sum_i p_i S( \rho^g_{i,A}),
\eea
which is inconsistent with the result (\ref{linearcondition}). Since the second order correction of the entanglement entropy $S_2$
is related to quadratic term  $\langle T_{\mu\nu}\rangle^2$. In general we have 
\bea\label{general2}
S(\rho_{c,A})= \sum_i p_i S( \rho^g_{i,A})+\frac{R^{2d}}{16\pi G}\big( c_2 E(X_{2,i}-E(X_{2,i}))^2+c_3 E(X_{3,i}-E(X_{3,i}))^2+c_4 E(X_{4,i}-E(X_{4,i}))^2   \big),
\eea
where $X_{1,i}:=\langle T_{00}\rangle_{\rho_i^g},X_{2,i}:=\langle T_{0k}\rangle_{\rho_i^g},X_{3,i}:=\langle T_{kl}\rangle_{\rho_i^g}$,$E(X)$ denotes the expectation value of the random variable $X$ with the probability $p_i$.
In the above derivation we only assume the state $\rho_c$ is still a geometric state. So the inconsistence between (\ref{linearcondition}) and (\ref{general2}) means this assumption is not right. We arrive at our result (\ref{rhocon}) in this section that the state $\rho_c$ cannot be a geometric state.
\subsubsection{2D CFTs}
For 2D CFTs  we can evaluate the R\' enyi or entanglement entropy of a short interval by using the operator product expansion (OPE) of twist operators in the n-fold  CFTs \cite{Headrick:2010zt},\cite{Cardy:2007mb}-\cite{Chen:2013dxa}, see also  a short review in \cite{Guo:2018pvi}. In this section we will make the calculation for 2D CFTs in the large central charge limit $c\to \infty$ , with the assumption that the vacuum family domains the contributions of the OPE of twist operators \cite{Headrick:2010zt}. \\
We consider the CFT lives on a cylinder with the spatial period $L$, the subsystem $A$ is the interval $[0,\ell]$ with $\ell\ll L$. The entanglement entropy of $A$ in the state $\rho$ up to $O(\ell^4)$ is given by 
\bea\label{entanglemententropy2D}
S_{A,\rho}=\frac{c}{6}\log \frac{\ell}{\epsilon}+a_T\langle T\rangle_\rho \ell^2+\frac{a_T}{2}\partial \langle T\rangle_\rho\ell^3 +(a_{TT}\langle T\rangle_\rho^2+\frac{3}{20}\partial^2\langle T\rangle_\rho)\ell^4+O(\ell^5),
\eea
where $a_T=-\frac{1}{6}, a_{TT}=-\frac{1}{30c}$.  we only consider the contributions from the holomorphic part of the vacuum conformal family.\\
For relative entropy we have following expansion,
\bea\label{relativeentropy2D}
S(\rho||\sigma)=-\ell^4 a_{TT}(\langle T\rangle_\rho-\langle T\rangle_\sigma)^2+O(\ell^5).
\eea
We can follow the same step as last section to study the convex combination of geometric states like (\ref{rhocon}). For simplicity we will use the same notation for the states. We have
\bea
\langle T\rangle_{\rho_c}=\sum_i p_i \langle T\rangle_{\rho_i^g}.
\eea 
 Let's consider the energy density of the state $\rho_{\rho^g_i}$ is of order $c$, i.e., $\langle T\rangle_{\rho_i^g}\sim O(c)$. This is a necessary condition to keep the entanglement entropy to be $O(c)$ or $O(1/G)$ by using the relation $c=\frac{3}{2G}$.  By using (\ref{relativeentropy2D}) we have
\bea
S(\rho^g_{i,A}||\rho_{0,A})=- \ell^4 a_{TT}(\langle  T\rangle_{\rho_i^g}-\langle T\rangle_{\rho_0})^2+O(\ell^5)\sim O(c),
\eea
where $\rho^g_{i,A}$,$\rho_{0,A}$ are reduced density matrix of the geometric state $\rho_i^g$ and vacuum $\rho_0$. We can use the bound of Holevo information  similar as the higher dimension case and find the entanglement entropy of $A$ in the state $\rho_c$ is 
\bea\label{EE2DHolevo}
S(\rho_{c,A})=\sum_i p_i S(\rho_{i,A}^g)+O(\sqrt{c}),
\eea
in the large $c$ limit. But by using (\ref{entanglemententropy2D}) we have 
\bea\label{EE2Dcon}
S(\rho_{c,A})= \sum_i p_i S(\rho_{i,A}^g) -a_{TT}E (X_i-E(X_i))^2\ell^4+O(\ell^5),
\eea
where $X_i:=\langle T \rangle_{\rho_i^g}$, $E(X_i)$  denotes the expectation value of the random variable $X_i$ with the probability $p_i$. In general the term of $O(\ell^4)$ in (\ref{EE2Dcon})  is of order $c$, which is inconsistent with (\ref{EE2DHolevo}). It seems the above derivation is independent with  the holographic results, but we use the assumption the entanglement entropy is of order $c$. In \cite{Guo:2018fnv} the authors show the necessary condition of geometric state is that the entanglement entropy is of order $c$. So our result shows the state $\rho_c$ cannot be a geometric state even though the entanglement entropy of $A$ in the state $\rho_c$ is of order $c$.\\
\subsection{Holevo information of thermal states for 2D CFTs}\label{Holevothermal}
In last section we derive an interesting property of geometric states by using the Holevo information. In QFTs given an ensemble of a state it is generally not an easy task to calculate Holevo information. An example is shown in  \cite{Guo:2018djz}. In this section we will briefly review the Holevo information in 2D CFTs and derive a result that is useful in next section. \\
Consider the canonical and microcanonical thermal state in 2D CFT.  The canonical thermal state with fixed temperature $\beta$ is 
\bea
\rho_{\beta}=\sum_i p_i |E_i\rangle \langle E_i|,
\eea
with $p_i=e^{-\beta E_i}/Z(\beta)$, $Z(\beta)=\sum_i e^{-\beta E_i}$. The microcanonical thermal state with fixed energy $E$ is
\bea
\rho_{\lambda}=\sum_i p_i |E_i\rangle \langle E_i|, \quad \text{with} \quad p_i=\frac{\delta(E_i-E)}{\Omega(E)},
\eea
where $\Omega(E)$ is the  density of state at the energy  $\frac{2\pi }{L}(E-\frac{c}{12})$. The subscript $\lambda:=\sqrt{\frac{c L^2}{12 E}}$ can be taken as the effective temperature of the microcanonical state.  
 For a subsystem $A$ with length $\ell$, we can calculate the Holevo bound of the state $\rho_A=\sum p_i \rho_{i,A}$ with $\rho_{i,A}=tr_{\bar A}|E_i\rangle \langle E_i|$. In \cite{Guo:2018djz} the authors have calculated the Holevo information of a short interval $\ell\ll L$ by using short interval expansion of twist operators. 
The Holevo information of canonical ensemble is 
\bea\label{canonicalholevo}
\chi_{\beta,A}(\ell)=\frac{2\pi^3\ell^4}{45\beta^3 L}-\frac{8\pi^4\ell^6(\pi c L+12\beta)}{945c\beta^5L^2}+...+O(\ell^{12}).
\eea
The results are valid for $\ell\ll \beta \ll L$. Here we only show the results up to $O(\ell^6)$, 
one can find the results up to $O(\ell^{10})$ in \cite{Guo:2018djz}. The result of microcanonical ensemble is 
\bea\label{microcanonicalholevo}
\chi_{\lambda,A}(\ell)=\frac{\pi^3\ell^4\left[\pi c L(I_3-I_1)+24\lambda I_2\right]}{540\lambda^4 L I_1}+...+O(\ell^{12}).
\eea
 $I_\nu$ is the modified Bessel function of the first kind with the argument $\frac{\pi c L}{3\lambda}$. In the limit $L\gg \lambda$, $I_\nu/I_1\to 1$ by using the asymptotic expansion of $I_\nu(z)\simeq \frac{e^z}{\sqrt{2\pi z}}$ for $z\gg 1$. In this limit the Holevo information of microcanonical ensemble (\ref{microcanonicalholevo}) will be same as the canonical one (\ref{canonicalholevo}) if $\beta=\lambda$. \\
We can see that all the terms of the Holevo information (\ref{canonicalholevo})(\ref{microcanonicalholevo}) are at the order $c^0$ which can be seen as quantum correction in the large $c$ limit. And all the terms can be seen as finite size correction with the power $1/L$ of the system. Therefore, one would expect $\lim_{L\to \infty} \chi_{\beta,A},\chi_{\lambda,A}\to 0$ up to $O(\ell^{10})$.  \\
To calculate the Holevo information (\ref{holevodefinition}) we need to evaluate the entanglement entropy of subsystem $A$ in the state $\rho_{\beta}$ and $|E_i\rangle$.  See the  Appendix.\ref{appendixholevo} for a short review on the calculation. For a short interval we can use the short interval expansion method to get the entanglement entropy.  In general, the difficulty is to calculate the average product of one-point functions, such as the following forms,
\bea\label{functionj}
\mathcal{G}^{\text{diag}}_{ \mathcal{X}_1 \mathcal{X}_2... \mathcal{X}_k}(p_i):=\sum_i p_i \langle \mathcal{X}_1\rangle_i \langle \mathcal{X}_2 \rangle_i ... \langle \mathcal{X}_k\rangle_i ,
\eea
 where $\mathcal{X}_i$ are the quasi-primary operators that appears in the OPE of twist operators, $\langle \mathcal{X}\rangle_i$ denotes the expectation value of operator $\mathcal{X}$ in the state $|E_i\rangle$. For example, for $k=2$, $\mathcal{X}_1=\mathcal{X}_2=T$ in the microcanonical ensemble, we have
 \bea
 \sum_i p_i \langle T\rangle_i^2 =\frac{c\pi^3(24\lambda I_2 +cL\pi I_3)}{36L\lambda^4 I_1}.
 \eea
We will denote the function (\ref{functionj}) as $\mathcal{G}^{\text{diag}}_{ \mathcal{X}_1 \mathcal{X}_2... \mathcal{X}_k}(\beta)$ for the canonical ensemble state and   $\mathcal{G}^{\text{diag}}_{ \mathcal{X}_1 \mathcal{X}_2... \mathcal{X}_k}(\lambda)$ for the microcanonical ensemble state. \\
 Let's first consider the microcanonical ensemble. In the limit $L\to \infty$ we have 
 \bea\label{holevo12}
 \chi_{\lambda,A}=\ell^{12}\left[a_{\mathcal{B}\mathcal{B}}\left(\langle \mathcal{B}\rangle^2_{\lambda}-\sum_i p_i \langle \mathcal{B}\rangle_i^2\right)+a_{\mathcal{D}\mathcal{D}}\left(\langle \mathcal{D}\rangle_{\lambda}^2-\sum_i p_i \langle \mathcal{D}\rangle_i^2\right)\right]+O(\ell^{14}),
 \eea 
 with $p_i=\frac{\delta(E_i-E)}{\Omega(E)}$, where the energy density $\Omega(E)$ is given by the Cardy formula
$\Omega(E)\simeq\lambda I_1(\frac{\pi c L}{3\lambda})$,
$a_{\mathcal{B}\mathcal{B}}$ and $a_{\mathcal{D}\mathcal{D}}$ are constant coefficients 
\bea
a_{\mathcal{B}\mathcal{B}}=-\frac{25}{123552 c (70 c+29)}, ~~
   a_{\mathcal{D}\mathcal{D}} = -\frac{70 c+29}{18018 c (2 c-1) (5 c+22) (7 c+68)}.
\eea
We show this formula in the Appendix.\ref{appendixholevo}, where we define the function
\bea
\mathcal{G}_{\mathcal{X}_1...\mathcal{X}_k}(\lambda):=\frac{1}{\Omega(E)}\sum_{i_1...i_k}\langle E_{i_1}|\mathcal{X}_1|E_{i_2}\rangle\langle E_{i_2}|\mathcal{X}_2|E_{i_3}\rangle...\langle E_{i_k}|\mathcal{X}_k|E_{i_1}\rangle\delta(E_{i_1}-E)...\delta(E_{i_k}-E).
\eea
For $k=2$ and $\mathcal{X}_1=\mathcal{X}_2=\mathcal{X}$ we have
\bea\label{functiong}
&&\mathcal{G}_{\mathcal{X}\mathcal{X}}(\lambda)=\frac{1}{\Omega(E)}\sum_{i,i'}\langle E_{i}|\mathcal{X}|E_{i'}\rangle\langle E_{i'}|\mathcal{X}|E_{i}\rangle\delta(E_{i}-E)\delta(E_{i'}-E)\nonumber \\
&&\phantom{\mathcal{G}_{\mathcal{X}\mathcal{X}}(E)}= \mathcal{G}^{\text{diag}}_{ \mathcal{X}\mathcal{X}}(\lambda)+\mathcal{G}^{\text{off-diag}}_{\mathcal{X}\mathcal{X}}(\lambda)
\eea
with
\bea 
\mathcal{G}^{\text{off-diag}}_{\mathcal{X}\mathcal{X}}(\lambda):=\frac{1}{\Omega(E)}\sum_{i'\ne i}\langle E_{i}|\mathcal{X}|E_{i'}\rangle\langle E_{i'}|\mathcal{X}|E_{i}\rangle\delta(E_{i}-E)\delta(E_{i'}-E)
\eea
From (\ref{holevo12}) we can see the calculation of Holevo informaiton at the order $\ell^{12}$ is associated with the first term of (\ref{functiong}).  The function $\mathcal{G}_{\mathcal{X}_1...\mathcal{X}_k}(\lambda)$  have a simple expression in the limit $L\to \infty$\cite{Guo:2018fye}, i.e.,
\bea\label{limitfunctiong}
\lim_{L\to \infty} \mathcal{G}_{\mathcal{X}_1...\mathcal{X}_k}(\lambda)=\langle \mathcal{X}_1\rangle_{\lambda}\langle \mathcal{X}_2\rangle_{\lambda}...\langle \mathcal{X}_k\rangle_{\lambda},
\eea
where the one-point function of the quasi-primary operator in the microcanonical ensemble state is defined as
\bea
\langle \mathcal{X}\rangle_\lambda :=\frac{1}{\Omega(E)}\sum_i \langle \mathcal{X}\rangle_i \delta(E_i-E).
\eea
Using (\ref{limitfunctiong}) and  taking (\ref{functiong}) into (\ref{holevo12}) , we have
\bea
\chi_{\lambda,A}=\ell^{12}\left[a_{\mathcal{B}\mathcal{B}}\left(\lim_{L\to\infty}\mathcal{G}^{\text{off-diag}}_{\mathcal{B}\mathcal{B}}(\lambda)\right)+a_{\mathcal{D}\mathcal{D}} \left(\lim_{L\to\infty}\mathcal{G}^{\text{off-diag}}_{\mathcal{D}\mathcal{D}}(\lambda)\right)\right]+O(\ell^{14}).
\eea
Note that the coefficients $a_{\mathcal{B}\mathcal{B}}$ and $a_{\mathcal{D}\mathcal{D}}$ are negative in the large $c$ limit.  The term $\mathcal{G}^{\text{off-diag}}_{\mathcal{X}\mathcal{X}}$ is always positive by definition. Therefore, we conclude the Holevo information must be vanishing at the order $O(\ell^{12})$, that is
\bea
\lim_{L\to \infty}\chi_{\lambda,A}=0+O(\ell^{{14}}).
\eea
We also get
\bea\label{constraintBD}
\lim_{L\to\infty}\mathcal{G}^{\text{off-diag}}_{\mathcal{B}\mathcal{B}}(\lambda)=0 \quad \text{and} \quad \lim_{L\to\infty}\mathcal{G}^{\text{off-diag}}_{\mathcal{D}\mathcal{D}}(\lambda) =0.
\eea
As we have noted in the Appendix.\ref{appendixholevo},  the state $E_i$ can be organized as the common eigenstates  of the zero mode of  $T$ and $\mathcal{A}$, which leads to 
\bea
\mathcal{G}^{\text{off-diag}}_{TT}(\lambda)=0\quad \text{and} \quad \mathcal{G}^{\text{off-diag}}_{\mathcal{A}\mathcal{A}}(\lambda)=0.
\eea 
But for $\mathcal{X}=\mathcal{B},\mathcal{D}$,  (\ref{constraintBD}) is a non-trivial result.  \\
The canonical ensemble thermal state is associated with the microcanonical ensemble thermal state by using the Laplace transformation. The Holevo information  $\chi_{\beta,A}$  in the limit $L\to\infty$ is same as   (\ref{holevo12}) by replacement $\lambda\to \beta$ and $p_i \to e^{-\beta E_i} /Z(\beta)$. We also have the following relation 
\bea\label{canmicrelation}
 \mathcal{G}^{\text{diag}}_{ \mathcal{X}_1 \mathcal{X}_2... \mathcal{X}_k}(\beta)=\frac{1}{Z(\beta)}\int dE e^{-\beta E}\Omega(E) \mathcal{G}^{\text{diag}}_{ \mathcal{X}_1 \mathcal{X}_2... \mathcal{X}_k}(\lambda).
\eea
Using the above results we can derive the Holevo informaion in the canonical ensemble state is vanishing at the order $O(\ell^{12})$ in the limit $L\to\infty$\footnote{A subtle point is the expression of $ \mathcal{G}^{\text{diag}}_{ \mathcal{X}_1 \mathcal{X}_2... \mathcal{X}_k}(\lambda)$  is available for large energy $E$.  We will comment on this point in next subsection.}. \\
It is not an easy work to generalize the results (\ref{holevo12}) to $O(\ell^{14})$. At present the short interval expansion of the entanglement entropy at order $\ell^{14}$ haven't been worked out as far as we know. For higher order it is expected the results will include more quasi-primary operators. It is not a practical solution to work order by order. If one could prove $\mathcal{G}^{\text{diag}}_{\mathcal{X}_1\mathcal{X}_2...\mathcal{X}_k}(\lambda)\to \langle \mathcal{X}_1\rangle_{\lambda}\langle \mathcal{X}_2\rangle_{\lambda}...\langle \mathcal{X}_k\rangle_{\lambda}$ for the general quasi-primary operators in the limit $L\to\infty$, we will have $\lim_{L\to\infty}\chi_{\lambda,A}=0$. Here we only leave this as a conjecture.

\subsection{Primary operator}\label{primaryoperator}
In last section we only consider the contribution to the entanglement entropy is from the quasi-primary operators in the vacuum family. So we get the constraints (\ref{constraintBD}), or equally 
\bea
\lim_{L\to \infty}\mathcal{G}^{\text{diag}}_{\mathcal{X}\mathcal{X}}(\lambda)=\langle \mathcal{X}\rangle^2_\lambda,
\eea
for $\mathcal{X}=\mathcal{B}$ or $\mathcal{D}$. In general, the one point function $\langle \mathcal{X}\rangle_\lambda$ is non-vanishing for the quasi-primary operators.   For a primary operator $O$ we have the one-point function in the microcanonical ensemble state 
\bea
\langle O\rangle_\lambda= \frac{1}{\Omega(E)}\sum_i \langle E_i |O|E_i\rangle \delta(E_i-E).
\eea
Using the modular covariance of torus one-point function,  one may obtain $\langle O\rangle_\lambda$, which is associated the lightest operator $\mathcal{Y}$ with non-vanishing three-point coefficient $\langle \mathcal{Y}|\mathcal{X}|\mathcal{Y}\rangle$\cite{Kraus:2016nwo}. Here we are only interested in the thermodynamic limit $L\to \infty$.  The one-point function $\langle O\rangle_\lambda$ is exponentially suppressed \cite{Kraus:2016nwo}. Therefore, we can take $\langle O\rangle_\lambda=0$ in this limit. \\
By using (\ref{limitfunctiong}) we have
\bea\label{OOmic}
\lim_{L\to \infty}\mathcal{G}_{OO}(\lambda)=\lim_{L\to \infty}\mathcal{G}^{\text{diag}}_{OO}(\lambda)+\lim_{L\to \infty}\mathcal{G}^{\text{off-diag}}_{OO}(\lambda)=0.
\eea
Since $\mathcal{G}^{\text{diag}}_{OO}(\lambda), \mathcal{G}^{\text{off-diag}}_{OO}(\lambda)\ge 0$ by definition, we conclude 
\bea\label{primaryconstraint}
\lim_{L\to \infty}\mathcal{G}^{\text{diag}}_{OO}(\lambda)=0,\quad \text{and} \quad \lim_{L\to \infty}\mathcal{G}^{\text{off-diag}}_{OO}(\lambda)=0.
\eea
From the relation (\ref{canmicrelation}) we can obtain $\mathcal{G}^{\text{diag}}_{OO}(\beta)$.  Though the intergral function is vanishing in the limit $L\to\infty$,  the intergration over $E$ may give a contribution like $L^\alpha$ ($\alpha>0$).  Following the method in \cite{Kraus:2016nwo} we can calculate the function $\mathcal{G}_{OO}(\lambda)$,
\bea\label{OOmicresult}
\mathcal{G}_{OO}(\lambda)\propto \frac{1}{L^{2\Delta_O}}(E-\frac{c}{12})^{\Delta_O-\frac{1}{2}}.
\eea
We take the limit $L\to \infty$ and keep the effective inverse temperature $\lambda\propto \sqrt{\frac{L^2}{E}}$ finite.  One can derive (\ref{OOmic}) by using the result (\ref{OOmicresult}).   The function for the canonical ensemble state is
\bea
\mathcal{G}_{OO}(\beta):=\frac{1}{Z(\beta)}\int dE e^{- \frac{2\pi \beta  }{L}(E-\frac{c}{12})} \Omega(E)\mathcal{G}_{OO}(\lambda).
\eea
The result of $\mathcal{G}_{OO}(\lambda)$ (\ref{OOmicresult}) is available only for the large $E $. But one can choose a truncation point $E_\Lambda$ of the intergration, below which (\ref{OOmicresult}) is not a good approximation,  The intergral is some constant,  further using $Z(\beta)\simeq e^{\frac{\pi L c}{6 \beta}}$,  we conclude that the contribution below the truncation point is vanishing . For the contribution above the truncation point we can evaluate it directly. The result is $\mathcal{G}_{OO}(\beta)\sim L^{-1/2}$ to 0 in the limit $L\to \infty$.

\section{Correlation and geometry}\label{correlationgeometry}
It is believed that the entanglement between underlying degree of freedom of quantum gravity plays an essential role in the emergence of connected spacetime\cite{VanRaamsdonk:2010pw}. If $\rho_1$ and $\rho_2$ are dual to two different spacetime regions, it is obvious that the product state $\rho_1\otimes \rho_2$ would represent two unrelated systems, the dual spacetime should not be connected. But if a bipartite state $\rho_{12}$ is an entangled state, the operations in subsystem $1$ will also effect the subsystem $2$. It seems more reasonable to take these kind of states as connected spacetime dual to $\rho_{12}$. In other words, we would expect the connected spacetime has non-vanishing correlations between different subregions. \\
However, along with this insight  there are still many interesting questions. If two subsystems only have classical correlation, is it possible the system has a geometric description? Entanglement is a phenomenon only in quantum mechanics, which has often been identified as quantum correlation. But it is not the only one, there exists quantum correlations for an unentangled states\cite{OZ}.  In general, for a state $\rho$ the classical correlations are as important as the quantum correlations. What is the role of the classical correlations if the state has a geometric dual? We only attempt to show some examples to catch a glimpse of the possible relation between correlation and geometry in the context of AdS/CFT. 
\subsection{Eternal black hole}
The eternal black hole is an important and popular example to show the relation between entanglement and geometry. The black hole in AdS$_3$ is dual to 2D CFT living on a spatial circle of size $L$ in the canonical ensemble thermal state with inverse temperature $\beta$. 
The black hole ensemble requires the condition $L\gg \beta$. Now consider the maximally extended black hole, i.e., the eternal black hole. The geometry has two asymptotically AdS regions. In the CFT side, this geometry is expected to be dual to the two copies of 2D CFTs with the wavefunction of the thermofield double state\cite{Maldacena:2001kr}
\bea\label{doublethermo}
|\Psi\rangle_\beta:= \frac{1}{\sqrt{Z(\beta)}}\sum_{i}e^{-\frac{\beta}{2}E_i}|E_i\rangle_1\otimes |E_i\rangle_2,
\eea
where $|E_i\rangle_{1(2)}$ are the eigenstates of the Hamiltonian $H_{1(2)}$ of the two CFTs, $Z(\beta)$ is the partition function $Z(\beta):=\sum_i e^{-\beta E_i}$.  The time evolution is under the Hamiltonian $H=H_1-H_2$. The reduced density matrix of system $1$ and $2$ is given by the canonical ensemble thermal state with inverse temperature $\beta$. So the correlation functions on one side CFT is
\bea
~_\beta\langle\Psi| O_1(x_1)O_1(x_2)...O_1(x_k)|\Psi\rangle_\beta=\frac{1}{Z(\beta)}tr\left[e^{-\beta H}O_1(x_1)O_1(x_2)...O_1(x_k)\right]:=\langle O_1(x_1)O_1(x_2)...O_1(x_k)\rangle_\beta.
\eea
The two-point correlator with operators inserted on the opposite boundaries is given by
\bea
&&~_\beta\langle \Psi|O_1(t_1,\phi_1)O_2(t_2,\phi_2)|\Psi\rangle_\beta\nonumber \\
&&=\langle O_1(t_1-i\frac{\beta}{2},\phi_1)O_2(t_2,\phi_2)\rangle_\beta \nonumber \\
&&=\left(\frac{2\pi}{\beta}\right)^{4\Delta_1}\frac{\delta_{O_1O_2}}{\left(\cosh(\frac{2\pi \phi_{12}}{\beta})+\cosh(\frac{2\pi t_{12}}{\beta})\right)^{2\Delta_1}},
\eea
where $\Delta_1$ is the scaling dimension of the operator $O_1$, $\phi_{12}=\phi_1-\phi_2$ and $t_{12}=t_1-t_2$. For $\phi_{12}=0$ and $t_{12}=0$ the correlator has no divergence, we have
\bea\label{twopointzero}
~_\beta\langle \Psi|O_1O_2|\Psi\rangle_\beta\propto \delta_{O_1O_2} \beta^{-4\Delta_1}.
\eea
Using the definition of thermofield double state (\ref{doublethermo}), one can expand the correlator (\ref{twopointzero}) as follows,
\bea
~_\beta\langle \Psi|O_1O_2|\Psi\rangle_\beta =\frac{1}{Z(\beta)}\sum_{i,j}e^{-\frac{\beta(E_i+E_j)}{2}}~_1\langle E_i|O_1|E_j\rangle_1~_2\langle E_i|O_2|E_j\rangle_2.
\eea
For simplicity we will take $O_1=O_2=O$ and drop the lable $1,2$ for the eigenstates. Replacing the sum with integral, we have
\bea\label{correlatordouble}
&&~_\beta\langle \Psi|OO|\Psi\rangle_\beta=\frac{1}{Z(\beta)}\sum_{i,j}e^{-\frac{\beta(E_i+E_j)}{2}}\langle E_i|O|E_j\rangle\langle E_i|O|E_j\rangle\nonumber \\
&&\phantom{~_\beta\langle \Psi|OO|\Psi\rangle_\beta}=\int dE \int dE' e^{\frac{-\beta(E+E')}{2}}\Omega(E)\Omega(E') \mathcal{J}_{OO}(E,E'), 
\eea
where  we define
\bea
\mathcal{J}_{OO}(E,E'):=\frac{1}{\Omega(E)\Omega(E')}\sum_{m,n}\delta(E_m-E)\delta(E_n-E')\langle E_m|O|E_n\rangle^2.
\eea
Using the correlator in thermofield double state, one can gain the function $\mathcal{J}_{OO}(E,E')$  \cite{Romero-Bermudez:2018dim}.
\subsection{States with less correlation}\label{eternallesscorrelation}
In this section we would like to study some states for which the correlation is different from the thermofield double state (\ref{doublethermo}).  We require the reduced density matrix of system $1$ and $2$ are still given by the canonical ensemble thermal state with inverse temperature $\beta$ or the microcanonical ensemble state with $\lambda$. This keeps the entanglement entropy between system $1$ and $2$ invariant.
The first example is the classical state 
\bea\label{rho1}
\rho_{I,\beta}=\frac{1}{Z(\beta)}\sum_i e^{-\beta E_i} |E_i\rangle_1~_1\langle E_i|\otimes |E_i\rangle_2~_2\langle E_i|.
\eea
Unlike the thermofield double state the state $\rho_{I,\beta}$ is a mixed state. Though the entanglement entropy between system $1$ and $2$ is same as the thermofield double state, this state has only classical correlation.\\
Consider the two-point correlator with operators inserted at $(t_1=0,\phi_1=0)$ and $(t_2=0,\phi_2=0)$, 
\bea\label{correlatorclassical}
&&tr(\rho_{c,\beta}OO)=\frac{1}{Z(\beta)}\sum_ie^{-\beta E_i} \langle O\rangle_i^2\nonumber \\
&&\phantom{tr(\rho_{I,\beta}OO)}=\frac{1}{Z(\beta)}\int dE e^{-\beta E} \Omega(E)\mathcal{G}^{\text{diag}}_{OO}(\lambda),
\eea
where we define $\langle O\rangle_i:=\langle E_i |O|E_i\rangle$ and drop the index of the system $1$ and $2$, in the second step we replace the sum  with intergral.  This correlator is the one that is associated with the calculation of Holevo information in the canonical ensemble state in section. . For $O=\mathcal{B}$ or $\mathcal{D}$ we have calculated
\bea
\sum_i p_i\langle \mathcal{B}\rangle_i^2\quad \text{and}\quad \sum_i p_i\langle \mathcal{D}\rangle_i^2,    
\eea
with $p_i=\frac{e^{-\beta E_i}}{Z(\beta)}$ in the thermodynamic limit. By using the Holevo information we get the costraint (\ref{constraintBD}), equally,
\bea
\lim_{L\to \infty} p_i \langle \mathcal{B}\rangle_i^2=\langle \mathcal{B}\rangle_\beta^2 \quad \text{and} \quad \lim_{L\to \infty} p_i \langle \mathcal{D}\rangle_i^2=\langle \mathcal{D}\rangle_\beta^2.
\eea
This means the connected correlators is vanishing for $O=\mB,\mD$. 
For general primary operators using the result in section. we have
\bea
tr(\rho_{c,\beta}OO)=0,
\eea
in the limit $L\to \infty$, that is the connected two-point correlation function is vanishing in the thermodynamic limit. But if the state $\rho_{I,\beta}$ has a connected geometry dual, there exists a shortest geodesic line connecting $(t_1=0,\phi_1=0)$ and $(t_2=0,\phi_2=0)$ with length $l$. The correlator is given by
\bea
\langle O(t_1=0,\phi_1=0) O(t_2=0,\phi_2=0)\rangle \sim e^{-m l},
\eea
where $m$ is the mass associated with the conformal dimension of operator $O$. The correlator in the state $\rho_{I,\beta}$ is inconsistent with the expectation from the holography. Roughly, we can say the correlation in the state $\rho_{I,\beta}$ is too weak to product the holographic result.  \\
Let's compare the correlator (\ref{correlatorclassical}) with the one in the thermofield double state (\ref{correlatordouble}). We can reformulate the correlator (\ref{correlatordouble}) as
\bea\label{sumdivide}
&&~_\beta\langle \Psi| OO|\Psi\rangle_\beta= \frac{1}{Z(\beta)}\sum_i e^{-\beta E_i} \langle O\rangle_i^2+\frac{1}{Z(\beta)}\sum_{i\ne i'}e^{-\beta E_i}\langle E_i|O|E_{i'}\rangle \langle E_{i'}|O|E_i\rangle\delta(E_i-E_{i'})\nonumber \\
&&\phantom{~_\beta\langle \Psi| OO|\Psi\rangle_\beta}+\frac{1}{Z(\beta)} \sum_{E_i\ne E_j}e^{-\frac{E_i+E_j}{2}}\langle E_i|O|E_j\rangle\langle E_j|O|E_i\rangle.
\eea
We divide the sum into three parts, the contributions from the two terms in first line of (\ref{sumdivide}) is vanishing (\ref{primaryconstraint}). While the  sum of the three point correlation coefficients $\langle E_i|O|E_{i'}\rangle$ between different energy level gives the main contributions to the correlation between the operators inserted at opposite boundary.\\
The second example is the state like the thermofield double state but with reduced density matrix being the microcanonical ensemble state,
\bea
|\Psi\rangle_{\lambda}:=\frac{1}{\sqrt{\Omega(E)}}\sum_{i}\delta(E_i-E)|E_i\rangle_1\otimes |E_i\rangle_2.
\eea
This can be seen as a maximally extangled state with the dimension $\Omega(E)$.  The reduced density matrix of subsystem $1$ or $2$ is the microcanonical ensemble state with energy $E$.  It has been shown the microcanonical ensemble state is almost undistinguishable from the canonical one for short interval if $\lambda=\beta$\cite{Guo:2018djz}. But for medium interval or long interval, one can find some probes to distinguish these two ensemble, such as R\'enyi entropy\cite{Guo:2018fye}\cite{Dong:2018lsk}.  The two-point correlator with $O_1=O_1=O$  is
\bea
~_\lambda\langle \Psi| O O|\Psi\rangle_\lambda=\frac{1}{\Omega(E)}\sum_{i,j} \langle E_j|O|E_i\rangle \langle E_i|O|E_j\rangle \delta(E_i-E)\delta(E_j-E)=\mathcal{G}_{OO}(\lambda).
\eea
We have shown in section.\ref{primaryoperator} the function $\mathcal{G}_{OO}(\lambda)$ is vanishing in the limit $L\to \infty$.
In fact the state $|\Psi\rangle_\lambda$ is only a block of the thermofield double state $|\Psi\rangle_{\beta}$ in the sense that
\bea
|\Psi\rangle_\beta=\frac{1}{\sqrt{Z(\beta)}}\sum_{E}  e^{-\frac{\beta E}{2}}\sqrt{\Omega(E)} |\Psi\rangle_{\lambda}.
\eea
Using the state $|\Psi\rangle_\lambda$ we can construct the third example, which is convex combination of the states $|\Psi\rangle_\lambda$
\bea
\rho_{II,\beta}:=\frac{1}{Z(\beta)}\sum_E e^{-\beta E} \Omega(E)|\Psi\rangle_{\lambda} ~_\lambda \langle \Psi|.
\eea 
We also have the two-point correlator $tr(\rho_{II,\beta}O_1 O_2)=0$ with $O_1=O_1=O$.  The classical state $\rho_{I,\beta}$ is a separable state, which can be seen as non-entangled state.  But the state $\rho_{II,\beta}$ is non-separable, which means there exists quantum correlation between $1$ and $2$. But our results in this section show the correlations in state $\rho_{I,\beta}$ , $\rho_{II,\beta}$ and $|\Psi\rangle_\lambda$ are too weak comparing with holographic expectation.

\subsection{Classification of quantum states}\label{classification}
The three examples $\rho_{I,\beta}$, $\rho_{II,\beta}$ and $|\Psi\rangle_\lambda$ all cannot produce enough correlation for the eternal black hole.  The physical intuition is that the strength of quantum correlation between  system $1$ and $2$  should be in the following order,
\bea
Q(|\Psi\rangle_\beta) \ge Q(\rho_{II,\beta})\ge Q(|\Psi\rangle_\lambda)\ge Q(\rho_{I,\beta}),
\eea
where $Q(\rho)$ denote the quantum correlation in the state $\rho$. For the mixed state the entanglement entropy is no longer a good measure of quantum correlation.  In this section we would like to review some measures to characterize the quantum and classical correlation in a given quantum state. 
\subsubsection{Correlation and operation}
In general, correlation functions are taken as the fundamental quantities in QFTs. In quantum information theory the operation meaning of the quantities are more interesting. But in QFTs the operations are generally not well defined. For a given pure state $|\phi\rangle$,  an operation by a local operator  $O(x)$ is
\bea
|\phi'\rangle := \mathcal{N}_{O} O(x)|\phi\rangle.
\eea
To make the state $|\phi'\rangle$ to be a well defined state, one needs some regularization, for example the regularization procedure in \cite{Calabrese:2005in}.  For our purposes it is enough to know the normalization constant $\mathcal{N}_{O}$ after regularization is a positive constant.  One could also define the locally excited state with two operators, 
\bea
|\phi''\rangle = \mathcal{N}_{OO}O(x_1)O(x_2)|\phi\rangle.
\eea
The normalization constant $\mathcal{N}_{OO}$ is associated with the UV cut-off $\epsilon$.In general it is almost independent with the state $|\phi\rangle$.
It is obvious the two-point correlator of $O(x_1)$ and $O(x_2)$ in the state $|\phi\rangle$ is associated with the fidelity between $|\phi''\rangle$ and $|\phi\rangle$, i.e.,
\bea
\langle \phi|O(x_1)O(x_2)|\phi\rangle=\frac{1}{\mathcal{N}_{OO}}\mathcal{F}(|\phi''\rangle,|\phi\rangle),
\eea
where $\mathcal{F}(|\phi_1\rangle,|\phi_2\rangle):= |\langle \phi_1|\phi_2\rangle|$ is the fidelity between two pure states. 
Here for simplicity we assume $O_{1(2)}$ are Hermition operators. Now the physical meaning of the two point correlator is clear. One makes the operation $O(x_1)$ and $O(x_2)$ at the separate point $x_1$ and $x_2$ in the system with the state $|\phi\rangle$. The fidelity actually reflects the disturbance of the operations on the state $|\phi\rangle$. \\
For the mixed state the explanation is not so obvious. But for any mixed state $\rho_{\text{mix}}$ we can always formulate it as the ensemble 
\bea
\rho_{\text{mix}}=\sum_i p_i |\phi\rangle_i ~_i\langle \phi|.
\eea
The decomposition is not unique, but we would like to use the similar idea to reformulate the two-point correlator as
\bea
tr(\rho_{\text{mix}}O(x_1)O(x_2))\propto \sum_i p_i \mathcal{F}(|\phi''\rangle_i,|\phi\rangle_i),
\eea
where $|\phi''\rangle_i:=\mathcal{N}_{OO} O(x_1)O(x_2)|\phi\rangle_i$. We assume the normalization constant is independent with the reference state $|\phi\rangle_i$. \\
Of course we don't expect the operation with $O(x_1)$ and $O(x_2)$ corresponds to some real physical measurement. The above only show the relation between correlation and operation. This also motivates us to define some measures to charaterize correlation by using quantum measurement. The positive-operator-valued measurement (POVM) is the most general quantum measurement on a given state $\rho$.  It is described by a set of positive operator $E_a=M_a^\dagger M_a$, the state after measurement is
\bea
\rho'=\sum_a M_a \rho M_a^\dagger.
\eea
The probability of the outcome $a$ is given by $p_a=tr (E_a\rho)$. The projective measurement with a set of orthogonal projections $\{\Pi_k\}$ is a special case of POVM.\\
For a bipartite state $\rho_{AB}$ we have a set of projective measurement $\Pi^A_k$ and $\Pi_k^B$ for the subsystem $A$ and $B$ respectively. The two subsystem $A$ and $B$ have no correlation if the state $\rho_{AB}=\rho_A\otimes \rho_B$. The total correlation in the bipartite state $\rho_{AB}$ is quantified by the mutual information 
\bea
I(\rho_{AB})=S(\rho_A)+S(\rho_B)-S(\rho_{AB})=S(\rho_B)-S(\rho_B|\rho_A),
\eea
where $S(\rho_B|\rho_A):= S(\rho_{AB})-S(\rho_A)$ is the conditional entropy, which may be negative for quantum theory. 
In general, $I(\rho_{AB})$ includes both the classical and quantum correlations.  The idea to describe the classical correlation is to introduce a classical-quantum version of conditional entropy. If one makes the measurement by the set of projections $\Pi_{k}^A$ on the subsystem $A$, the state of $B$ after measurement is 
\bea
\rho'_{B}=\sum_k \Pi_k^A\rho_{AB}\Pi_k^A=\sum_k p_k \rho_{B}^k,
\eea
with $p_k=tr(\Pi_k^A \rho_{AB}\Pi_k^A)$ and
\bea
\rho_{B}^k=\frac{tr_A\Pi_k^A \rho_{AB}\Pi_k^A}{p_k}.
\eea
The new conditional entropy under the measurement is 
\bea
S(B|\Pi_k):= \sum_k p_k S(\rho_B^k),
\eea
where $S(\rho_k^B)$ can be seen as the missing information of the system $B$. To define the classical correlation  independent with special measurement, one can maximize over all the measurement.  
\bea
\mathcal{C}(B|A):= \text{max}_{\{\Pi_k\}}\left[S(B)-S(B|\Pi_k)\right].
\eea
The quantum correlation 
, denoted by $\mathcal{Q}(\rho_{AB})$, is given by the difference between the total correlation and the classical one, that is
\bea
\mathcal{Q}(B|A):=I(A,B)-\mathcal{C}(B|A).
\eea
$\mathcal{Q}$ is the quantum discord of the state $\rho_{AB}$\cite{OZ}\cite{LV}. If $\rho_{AB}$ is pure, $\mathcal{Q}(B|A)=S(\rho_{A})$.\\
By definition the quantity $Q(B|A)\ge 0$ for any state $\rho_{AB}$.  The quantum discord is not symmetric, i.e., $\mathcal{Q}(B|A)\ne \mathcal{Q}(A|B)$. The necessary and sufficient condition of the zero quantum discord $\mathcal{Q}(B|A)$ is \cite{OZ}
\bea\label{sufficientcondition}
\rho^{0}_{AB}=\sum_k p_k \Pi_k\otimes \rho_{B,k}.
\eea
Such states are called classical-quantum state, we denote the set of this kind of states as $\mathcal{Q}_0$. From this we can see a necessary condition for zero discord $\mathcal{Q}(B|A)$ is
\bea
[\rho^0_{AB},\rho^0_A]=0,
\eea
where $\rho^0_A=tr_B\rho^0_{AB}$\cite{Ferraro2010}.
\subsubsection{Classification of states by correlation}
With the quantum discord to quantify the correlation in a given bipartite state $\rho_{AB}$, one could roughly classify the quantum states into different sets.  The set zero discord state $\mathcal{Q}_0$ is the one with no quantum correlation.   A subset of  $\mathcal{Q}_0$ is  the classical state that satisfies 
\bea\label{classicalstate}
\Pi (\rho_{AB})=\rho_{AB},
\eea
with
\bea
\Pi(\rho_{AB}):=\sum_{ij} \Pi^A_i \Pi^B_j \rho_{AB}\Pi^A_i \Pi^B_j.
\eea
The subset is denoted by $\mathcal{C}_0$. We call these states classical states because they remain unchanged under the measurement.  \\
As proved in \cite{Luo08}, if a state $\rho_{AB}$ is classical (\ref{classicalstate}) with respect to the measurement $\Pi_i^A\otimes \Pi_j^B$, then   $\{\Pi_i^A\otimes \Pi_j^B\}$, $\{\Pi_i^A\}$ and $\{\Pi_j^B\}$ must be the eigestates of $\rho_{AB}$, its reduced density matrix $\rho_{A}$ and $\rho_{B}$, respectively.  Moreover, any classical state $\rho_{AB}\in \mathcal{C}_0$ can be represented as 
\bea\label{classicalstate}
\chi_{AB}=\sum_{ij}p_{ij}\Pi_{i}^A\otimes \Pi_j^B,
\eea
where $\{\Pi_i^A\}$ and $\{\Pi_j^B\}$ are the eigenstates of $\rho_{A}$ and $\rho_{B}$, $p_{ij}$ is the probability. These states can be identified to a system with the classical probability  $p_{ij}$. This is also the reason to call them classical states.  The state $\rho_{I,\beta}$  (\ref{rho1}) is a classical state.  \\
A larger set comparing with $\mathcal{Q}_0$ is the set of separable states $\mathcal{S}$, which includes the states of the form
\bea\label{separable}
\sigma_{AB}=\sum_i p_i \rho_i^A\otimes \rho_i^B, 
\eea
where $\rho_i^A$ and $\rho_i^B$ are states in $A$ and $B$. The separable state can be constructed by convex combination of product state.  $\mathcal{S}$ is a convex set.   Using the Bell inequality as an indicator for the existence of quantum correlaiton, the separable states should belong to the ``classical'' state, i.e., the Bell inequality is satisfied for the separable state\cite{Werner}.  However,  it has been shown in \cite{OZ} some separable states could have quantum correlation, that is with non-zero discord.  It is expected quantum discord $\mathcal{Q}$ is a more general quantity to capture quantum correlation. 
Finally, the states that cannot be represented as the separable form (\ref{separable}) are entangled states.

\subsubsection{Geometric measure of correlation}
Evaluation of quantum discord in general is difficult because of  the maximization procedure  over all the measurements. Moreover, for QFTs the definition of quantum measurement is still not clear.  To quantify the quantum and classical correlation one can  introduce a measure based on the distance between density matrices. A geometric measure is defined as\cite{Dakic2010}
\bea
\bar{\mathcal{Q}}^{(2)}(\rho_{AB})=\mathop{\text{min}}\limits_{\rho^0_{AB} \in \mathcal{Q}_0} ||\rho_{AB}-\rho^0_{AB}||^2,
\eea
where $||\rho||^2:=tr \rho^2$ is the Hilbert-Schmidt distance. One could also use other distance measure such as the Bures distance, trace distance or even the relative entropy\footnote{For the definition based on relative entropy\cite{Modi2009}, there is additivity relations between different correlations. But this nice property is not right in other cases.}.  One could refer to \cite{Modi2011} for a review on this topic.  For our purpose it is more useful to define a geometric measure with the minimization over the set of classical state $\mathcal{C}_0$, that is
\bea
\mathcal{Q}^{(2)}(\rho_{AB}):=\mathop{\text{min}}\limits_{\chi_{AB} \in \mathcal{C}_0} \frac{ ||\rho_{AB}-\chi_{AB}||^2}{||\rho_{AB}||^2},
\eea
Assume $\chi_{AB}^\rho$ is the closest classical state to $\rho_{AB}$.
The classical correlation can be quantified by the distance between the $\chi_{AB}^\rho$ and the product state $\pi^\rho_{AB}:=\pi^\rho_{A}\otimes \pi^\rho_B$ with $\pi^\rho_{A(B)}:=tr_{A(B)}\pi_{AB}^\rho$. The geometric measure of classical correlation $\mathcal{C}^{(2)}(\rho_{AB})$ is
\bea\label{classicalmeasure}
\mathcal{C}^{(2)}(\rho_{AB}):=\frac{||\chi^\rho_{AB}-\pi^\rho_{AB}||^2}{||\chi_{AB}||^2}.
\eea
In the above definition we introduce a normalization of quantum and classical correlation  for the state $\rho_{AB}$ and $\chi_{AB}$. 

\subsubsection{Reeh-Schlieder property and quantum correlation in QFTs}\label{RScorrelation}
We make a lot of discussions on the definition of quantum and classical correlation. Our motivation is to use them to study the correlations in states of QFTs.  \\
Consider two subsystem $A$ and $B$ that are spacelike. For simplicity, let's choose them on a time slice $t=0$ in the vacuum state, the minimal distance between them is $d$. The correlation between $A$ and $B$ is associated with the distance $d$. We can see this by considering the two-point correlator $\langle O_A O_B\rangle$ with $O_{A(B)}$ being the operators located in the region $A(B)$. Generally, the two-point correlator is dependent with the distance $d$ and non-vanishing even for large $d$\footnote{We say $d$ is large or small by comparing with some scale of the theory. For example, if the theory has a mass scale $m$, we say $d$ is large for $d\gg 1/m$. }.\\
We would like to show the reduced density matrix $\rho_{AB}$ cannot belong to the set $\mathcal{Q}_0$ even for very large $d$. 
Let's consider the vacuum state, which satisfies the Reeh-Schlieder property. To make clear what is the Reeh-Schlieder property, we need some basic elements of QFTs in the framework of algebra. One may refer to the book \cite{Haag}\cite{Terhal} or recent review \cite{Witten:2018lha}. \\
The starting point of algebraic QFTs is that any regions of the system, say $A$, can be associated with observable algebra $\mathcal{U}(A)$. The algebras satisfy the assumptions: \\
~\\
\quad \textit{Isotony}: $A_1\subset A_2$ $\Longrightarrow$ $\mathcal{U}(A_1)\subset \mathcal{U}(A_2)$.\\
\quad \textit{microcausality}: $A$ and $B$ are spacelike $\Longrightarrow$ $[\mathcal{U}(A),\mathcal{U}(B)]=0$.\\
~\\
The Reeh-Schlieder property of vacuum state is  summarized as follows. \\
~\\
\quad \textit{ For any given state $|\psi\rangle$ and any subregion $A$, it is always possible to find an operator $O(\psi)\in \mathcal{U}(A)$ such that the distance between $|\psi\rangle$ and $O(\psi)|0\rangle$ is almost vanishing, i.e.,
\bea
|\psi\rangle \overset{\epsilon}{\simeq} O(\psi)|0\rangle,
\eea
where we define $|\psi\rangle_1 \overset{\epsilon}{\simeq} |\psi\rangle_2$ if $||\psi\rangle-O(\psi)|0\rangle| <\epsilon$
for any positive constant $\epsilon$.}\\
~\\
This property is also true for other low energy excited states\cite{Witten:2018lha}, for example the thermal state.  \\
Consider two spacelike subregion $A$ and $B$. By using the Reeh-Schlieder property for any $O_{A}\in \mathcal{U}(A)$ there exists an operator $O_B\in\mathcal{U}(B)$ such that
\bea
O_A|0\rangle \overset{\epsilon}{\simeq} O_B |0\rangle.
\eea
In the following we will just take the notation $\overset{\epsilon}{\simeq}$ as $=$ for simplicity.
The above equation is equal to 
\bea\label{RS1}
O_A \rho_{AB} O_A = O_B\rho_{AB} O_B,
\eea
where we assume $O_{A(B)}$ is Hermitian.  If $\rho_{AB}\in \mathcal{Q}_0$, we have (\ref{sufficientcondition})
\bea
[\rho_{AB},\rho_{A}]=\rho_{AB}\rho_A-\rho_A\rho_{AB}=0.
\eea
Using (\ref{RS1}) and  the microcausality property we obtain
\bea
O_B \rho_{AB}O_{B} \rho_A-\rho_{A} O_B \rho_{AB} O_B=O_A\rho_{AB}O_A\rho_A-\rho_AO_A\rho_{AB} O_A=0.
\eea
Tracing over the degree of freedom of $B$ in above equation, we have
\bea
(O_A\rho_A)^2-(\rho_AO_A)^2=0.
\eea
This leads to $[\rho_A,O_A]=0$ or $\{\rho_A,O_A\}=0$ for \textit{any} operator $O_A\in \mathcal{U}(A)$. If $O_A$ is an operator commutating with $\rho_A$, $\{\rho_A,O_A\}\ne 0$. Therefore, the result is 
\bea
[\rho_A,O_A]=0,
\eea
for any $O_A$. This means $\rho_A$ should be a constant which is of course not true for any subsystem $A$ in the vacuum state. 
The assumption $\rho_{AB}\in \mathcal{Q}_0$ is not right. \\
Actually, we can further show the state $\rho_{AB} $ cannot be in the set of separable states $\mathcal{S}$.  This  also follows the Reeh-Schlieder property of the vacuum state\cite{Verch:2004vj}.  A popular criterion for separability of a bipartite state $\rho_{AB}$  is by the positivity of partial transpose (ppt). Given the orthonormal bases $\{|\lambda_i\rangle_A\}$ for the subsystem $A$ and $\{|\lambda_i\rangle_B\}$for $B$, the transpose of $\rho_{AB}$ with respect to $A$ is defined as 
\bea
~_A\langle \lambda_i| ~_B\langle \lambda_j| \rho^{T_A}_{AB}|\lambda_m\rangle_{A}|\lambda_n\rangle_B=~_A\langle \lambda_m| ~_B\langle \lambda_j| \rho_{AB}|\lambda_i\rangle_{A}|\lambda_n\rangle_B.
\eea
A necessary condition for separability of $\rho_{AB}$ is the positivity of the transpose matrix $\rho^{T_A}_{AB}$, i.e., $\rho^{T_A}_{AB}\ge 0$.  In \cite{Verch:2004vj} the authors show the Reeh-Schlieder property would lead to the non-positivity of  
transpose matrix.  Therefore, the state $\rho_{AB}$ cannot be separable. One can also get the conclusion by using the violation of Bell inequality for any spacelike separate region  in QFTs\cite{Summers:1985}\cite{Summers:1987fn}\cite{Summers:1987ze}.  It is a general phenomenon that there exist entanglement between two  arbitrary spacelike region in the state satisfying Reeh-Schlieder property.

\subsection{Quantum correlation in geometric state}
\subsubsection{Geometric measure}
We would like to show two examples that we can calculate geometric measure of quantum correlation $\mathcal{Q}^{(2)}$ in QFTs. The first example is the thermofield double state $|\Psi\rangle_{\beta}$ (\ref{doublethermo}) . The reduced density matrices of  system $1$ and $2$ is given by the canonical ensemble state. The eigenstates of the them is $\{|E_i\rangle_1 \}$ and $\{ |E_j\rangle_2 \}$.  The general classical state is 
\bea
\chi_{12}=\sum_{ij}q_{ij} |E_i\rangle_1 ~_1\langle E_i|\otimes  |E_j\rangle_2 ~_2\langle E_j|,
\eea
where $0\le q_{ij}\le 1$.  With some calculations we have
\bea
&&|||\Psi\rangle_\beta~_\beta \langle\Psi|-\chi_{12}||^2=1 -2 ~_\beta\langle \Psi| \chi_{12} |\Psi\rangle_\beta +tr \chi^2_{12}\nonumber \\
&&\phantom{|||\Psi\rangle_\beta~_\beta \langle\Psi|-\chi_{12}||^2}=1-2 \sum_{i}q_{ii}p_i^2+\sum_{ij}q_{ij}^2\nonumber \\
&&\phantom{|||\Psi\rangle_\beta~_\beta \langle\Psi|-\chi_{12}||^2}=1-\sum_i p_i^4+\sum_i (p_i^2-q_{ii})^2+\sum_{i\ne j}q_{ij}^2. 
\eea
where $p_i=e^{-\frac{\beta E_i}{2}}/\sqrt{Z(\beta)}$. The minimal value of the above expression is $1-\sum_i p_i^4$ with $q_{ii}=p_i^2$ and $q_{ij}=0$. The closest classical state $\chi^\rho_{12}$ is the classical state $\rho_{I,\beta}$ (\ref{rho1}) that we discuss in section.\ref{eternallesscorrelation} .\\
The geometric  quantum correlation is 
\bea
\mathcal{Q}^{(2)}(|\Psi\rangle_\beta)=1-\frac{\sum_i e^{-2\beta E_i}}{Z^2(\beta)}=1-\frac{Z(2\beta)}{Z^2(\beta)}.
\eea
For the 2D CFT, the partition funciton $Z(\beta)=e^{\frac{\pi L c}{6\beta}}$ in the high temperature limit $L\gg \beta$. 
We have
\bea
\mathcal{Q}^{(2)}(|\Psi\rangle_\beta)=1-e^{-\frac{\pi L c}{4\beta}}.
\eea
 For the fixed $L/\beta$  the quantum correlation will approach to   $1$  in the semi-classical limit $c\to \infty$.  By the definition of geometric measure of classical correlation,  we have
\bea
\mathcal{C}^{(2)}(|\Psi\rangle_\beta)=\frac{\sum_{i} q_{ii}^2-2\sum_i q_{ii}^3+(\sum_iq_{ii}^2)^2}{{\sum_i q^2_{ii}}}=1-\frac{Z(3\beta)}{Z(2\beta)Z(\beta)}+\frac{Z(2\beta)}{Z^2(\beta)}.
\eea 
For 2D CFTs we have $\mathcal{C}^{(2)}(\Psi\rangle_\beta) \simeq 1+e^{-\frac{\pi L c}{4\beta}}$. In the large $c$ limit we find the sum of the quantum and classical correlation is near $2$. \\
The second example is the pure  geometric state $|\Psi\rangle_g$. We will consider the correlation between $A$ and its complementary $\bar A$. The Schmidt decomposition of the pure state $|\Psi\rangle_g$ is
\bea
|\Psi\rangle_g= \sum_i \sqrt{\lambda_i} |\lambda\rangle_i \otimes |\bar\lambda\rangle_i,
\eea
where $\{|\lambda\rangle_i\}$ and $\{ |\bar \lambda\rangle_i\}$ are the eigenstates of $\rho_{A}$ and $\rho_{\bar A}$, respectively.  We can repeat the similar steps as the calculations of the state $|\Psi\rangle_\beta$ and get the quantum correlation
\bea\label{purequantumcorrelation}
\mathcal{Q}^{(2)}(|\Psi\rangle_g)=1 -e^{-S^{(2)}(\rho_A)},
\eea
where $S^{(2)}(\rho_A)$ is the R\'enyi entropy  $S^{(n)}(\rho_A):=\frac{\log \rho_A^n}{1-n}$  with $n=2$.  If the state $|\Psi\rangle_g$ has a geometric dual,  one could use the holographic proposal of R\'enyi entropy \cite{Dong:2016fnf}. Generally,  it is of the form
\bea
S^{(2)}(\rho_A)=\frac{1}{G} s_2,
\eea
where $s_2>0$ is associated with a geometric quantity of the bulk manifold. In the semiclassical limit $G\to 0$, we find $\mathcal{Q}^{(2)}(\rho_A)\simeq 1$. \\ 
From (\ref{purequantumcorrelation}) we see that the quantum correlation of a pure state is related to the R\'enyi entropy with $n=2$. This result is not so interesting, since the R\'enyi entropy is a good measure of entanglement for a pure state. Unfortunately, we have no good examples of mixed states in QFTs that can be evaluated at present. 

\subsection{Correlation in geometric state}\label{final}
\subsubsection{Subsystems with large distance}
Consider two spacelike regions $A$ and $B$ with a minimal distance $d$. We have shown in section.\ref{RScorrelation} the entanglement or quantum correlation between $A$ and $B$ is non-vanishing for any states satisfying Reeh-Schlieder property, such as the vacuum. The behavior of the strength of quantum correlation with respect to different $d$ is an important indicator for the state and the underlying theory.  We are interested in the theory with gravity dual. Assume a state $\rho$ has a geometry description at the semiclassical limit $G\to 0$. By using the RT formula, the mutual information of $\rho_{AB}:=tr_{\overline{AB}}\rho$ is vanishing  for large enough distance $d$ in the limit $G\to 0$. That is
\bea
I(\rho_{AB})=0+O(G^0).
\eea   
Since $I(\rho_{AB})= S(\rho_{AB}|\rho_{A}\otimes \rho_B)$, we have
\bea\label{quasiproduct}
\rho_{AB}= \rho_A\otimes \rho_{B}+\delta\rho_{AB},
\eea
with $\delta\rho_{AB}$ being small in the sense that it can be taken as a perturbation in the limit $G\to 0$. From the field theory calculation we know $I(\rho_{AB})\sim O(G^0)$ in the vacuum  state for large enough $d$. Therefore, for large enough $d$ we expect $\rho_{AB}$ can be roughly taken as product state $\rho_{A}\otimes \rho_B$ in $G\to 0$. This means $A$ and $B$ almost lose correlation in this case. \\
By using this fact we can provide a more simpler understanding on the conclusion in section.\ref{convexgeometry}. From (\ref{rhocon}) we have
\bea
\rho_{c,AB}=\sum_i p_i \rho^g_{i,AB}, 
\eea
where $\rho_{i,AB}:=tr_{\overline{AB}}\rho^g_{i}$ and $\rho_{c,AB}:=tr_{\overline{AB}}\rho_c$. For any geometric states $\rho_i^g$ we expect $\rho_{i,AB}^g\simeq \rho_{i,A}^g\otimes \rho_{i,B}^g$ for large enough $d$. But $\rho_{c,AB}$ is a separable state 
\bea
\rho_{c,AB}\simeq \sum_i p_i \rho_{i,A}^g\otimes \rho_{i,B}^g,
\eea
 instead of a product state. This means   there exists correlation between $A$ and $B$ in the state $\rho_{c, AB}$, in general the quantum correlation is also non-vanishing.  
\subsubsection{Quasi-product states}
By the clustering property of vacuum state we expect the reduced density matrix $\rho_{AB}$ should be like (\ref{quasiproduct}) and $\delta \rho_{AB}$ is a function of the distance $d$. But for the theory with gravity dual $\delta \rho_{AB}$ should also be associated with the central charge $c\sim 1/G$. A more important feature is that there exists an phase transition for the mutual information $I(\rho_{AB})$. Consider  2D CFTs in the vacuum state, $A$ and  $B$ are two interval with length $l_A$ and $l_B$ and distance $l$ as shown in Fig. .  We have
\bea\label{phasetransition}
I(\rho_{AB})=\begin{cases} 0, & l>l_c, \\  \frac{c}{3}\log \left(\frac{l \left(l_A+l_B+l\right)}{l_A l_B}\right), &l\le l_c,\end{cases}
\eea
where $l_c:= \frac{1}{2} \left(\sqrt{l_A^2+l_B^2+6 l_A l_B}-l_A-l_B\right)$, we only keep the leading order of $c$.  The dependence of $\delta \rho_{AB}$ on $c$ and $l$ is the key to understand this feature of theories with holographic dual. \\
Let's first consider the entanglement entropy of one interval, e.g., the subsystem $A$.  By using the R\'enyi entropy one can derive the distribution of eigenvalues of $\rho_A$. The distribution $P(\lambda):=\sum_i \delta(\lambda_i-\lambda)$ is given by
\bea
P(\lambda)=\delta(\lambda_{m}-\lambda)+\frac{b \theta(\lambda_m-\lambda)}{\lambda\sqrt{b\log \lambda_m/\lambda}}I_1(2\sqrt{b\log(\lambda_m/\lambda)}),
\eea
where $\lambda_m$ is the maximum eigenvalue, $b=-\log \lambda_m$, $I_1(x)$ is the modified Bessel function of the first kind. 
$\lambda_m$ is associated with the central charge $c$ and size $l_A$. For the vacuum state $\lambda_m =e^{-\frac{c}{6}\log l_A/\epsilon}$, which is exponentially suppressed in large $c$ limit.  $P(\lambda)$ satisfies the normalization  $\int_{0}^{\lambda_m} \lambda P(\lambda)d\lambda=1$.  Taking $\lambda=\lambda_m e^{-b y^2}$ the integral of  the normalization is
\bea
\int_0^{+\infty}  p_n(y)dy=1 \quad \text{with}\quad p_n(y)=\lambda_m \delta(y)+\frac{b e^{-b (y - 1)^2}}{\sqrt{\pi}\sqrt{b y}} .
\eea
In the limit $c\to \infty$ or equally $b\to \infty$ we have
\bea\label{distributionnorm}
p_n(y)\simeq \frac{1}{\sqrt{y}}\delta(y-1),
\eea
where we use $\delta(x)=\lim_{\epsilon\to0} \frac{1}{\sqrt{2\pi \epsilon}}e^{-\frac{x^2}{2\epsilon}}$.  In the above equation we ignore the term $\lambda_m \delta(y)$ which is exponentially suppressed $e^{-c}$ in large $c$ limit.
One can also check the entanglement entropy $S_A=-\int_0^{\lambda_m} \lambda \log \lambda P(\lambda)d\lambda=-2\log \lambda_m$. Taking $\lambda=\lambda_m e^{-b z^2}$ the integral of $S_A$ becomes
\bea
S_{A}=\int_0^{+\infty}p_e(z)dz\quad \text{with}\quad p_e(z) :=-\lambda_m \log \lambda_m \delta(z)+ 2 b^2 \left(z^2+1\right) e^{-b \left(z^2+1\right)} I_1 (2 b z). 
\eea 
We are interested in the large $c$ limit.  $p_e(z)$ can be approximated by
\bea
p_e(z) \simeq -\lambda_m \log \lambda_m \delta(z) +\frac{b^2 \left(z^2+1\right) e^{-b (z-1)^2}}{\sqrt{\pi } \sqrt{b z}},
\eea 
where we use the approximation $I_1(x)\simeq \frac{e^x}{\sqrt{2\pi x}}$ for $x\gg 1$. The contribution of the delta function in $p_e(z)$ can be ignored, which is $O(e^{-c})$.  In the limit $c\to \infty$ or $b\to \infty$ we can reformulate $p(z)$ more simpler as
\bea\label{distributionentropy}
p_e(z)\simeq \frac{b \left(z^2+1\right)}{\sqrt{z}}\delta(z-1).
\eea
 This means the main contributions to the entanglement entropy in holographic theory is near the eigenvalue $\lambda_0:= e^{-2b}$. \\
The distribution of the  eigenvalues $\lambda_A \lambda_B$ of the product state $\rho_A\otimes \rho_B$ is $P(\lambda_A)P(\lambda_B)$.  For the state $\rho_{AB}$ (\ref{quasiproduct}) the eigenvalues should be $\lambda_A\lambda_B+\delta(\lambda_A,\lambda_B)$ with $\delta(\lambda_A,\lambda_B)\ll \lambda_A\lambda_B$.  It is expected the perturbation $\delta(\lambda_A,\lambda_B)$ should depend on the distance $l$ and central charge $c$.  We can calculate the entanglement entropy 
\bea
&&S(\rho_{AB}):=-tr \rho_{AB}\log \rho_{AB}=-\sum_{ij} [\lambda_{A,i}\lambda_{A,j}+\delta(\lambda_{A,i},\lambda_{A,j})]\log [\lambda_{A,i}\lambda_{A,j}+\delta(\lambda_{A,i},\lambda_{A,j})] 
\nonumber \\
&&\phantom{S(\rho_{AB}):=-tr \rho_{AB}\log \rho_{AB}}\simeq -\sum_{i,j}\lambda_{A,i}\lambda_{B,j}\log(\lambda_{A,i}\lambda_{B,j})-\delta(\lambda_{A,i},\lambda_{B,i})\log(\lambda_{A,i}\lambda_{B,i}) \nonumber \\ 
&&\phantom{S(\rho_{AB}):=-tr \rho_{AB}\log \rho_{AB}}=S(\rho_A)+S(\rho_B)-\sum_{ij}\delta(\lambda_{A,i},\lambda_{B,i})\log(\lambda_{A,i}\lambda_{B,i})
\eea
where in the second step we only keep the leading order contribution. For simplifying the notations we express the formula by  discrete sum.  Replacing the sum with integral we have
\bea
&&I(\rho_{AB})=-\int_0^{\lambda_{A,m}} d\lambda_A  \int_0^{\lambda_{B,m}} d\lambda_B \frac{\delta(\lambda_A,\lambda_B)}{\lambda_A\lambda_B} \left[\lambda_A\log \lambda_A P(\lambda_A) \lambda_BP(\lambda_B)+\lambda_B\log \lambda_B P(\lambda_B) \lambda_AP(\lambda_A)\right]\nonumber \\
&&\phantom{I(\rho_{AB})}=\int_0^{+\infty}dz_A \int_0^{+\infty}dz_B   \frac{\delta(\lambda_A,\lambda_B)}{\lambda_A\lambda_B} \left[p_e(z_A)p_n(z_B)+p_n(z_A)p_e(z_B)\right],
\eea 
where in the second step we use $\lambda_A=\lambda_{A,m}e^{-b_Az_A^2 }$ and $\lambda_B=\lambda_{B,m}e^{-b_Bz_B^2 }$, $b_{A(B)}=-\log \lambda_{m,A(B)}=\frac{c}{6}\log l_{A(B)}/\epsilon$. By using (\ref{distributionnorm}) and (\ref{distributionentropy}) we have
\bea\label{mutualspectrum}
I(\rho_{AB})= 2(b_A+b_B) \frac{\delta(\lambda_{A,0},\lambda_{B,0})}{\lambda_{A,0}\lambda_{B,0}},
\eea
with $\lambda_{A(B),0}:= \lambda_{A(B),m}e^{-b_{A(B)}}$.  Now we can make clear what is the role of $c$ in the perturbative expansion of $\rho_{AB}$ (\ref{quasiproduct}).   For $l> l_c$  from (\ref{phasetransition}) we expect $\overline \delta_0 :=\frac{\delta(\lambda_{A,0},\lambda_{B,0})}{\lambda_{A,0}\lambda_{B,0}}$ should be of order $\frac{1}{c}$. The perturbation calculation is reliable in this case.  For $l\le l_c$, $\bar \delta_0$ is of order $c^0$, the perturbation calculation may broke down.  It is still unclear why the transition occurs at the point $l=l_c$ because the lack of the dependence of $\bar \delta_0$ on $l$. This is a problem of eigenvalues perturbation, but we almost know nothing on the form of $\delta\rho_{AB}$ .   By using (\ref{mutualspectrum}) and (\ref{phasetransition}) it seems we can gain some information on the spectrum of perturbation matrix $\delta\rho_{AB}$. 

\subsubsection{More on  $A$ and $B$}
\begin{figure}[H]
\centering 
\includegraphics[trim = 0mm 0mm 10mm 0mm, clip=true,width=10.0cm]{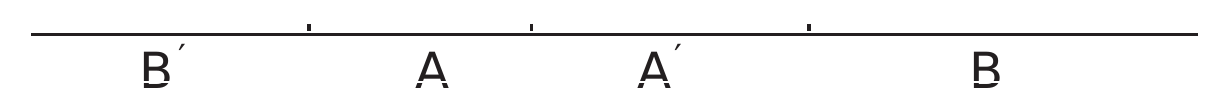}
\caption{Subsystems $A$, $B$ and $A'$, $B'$.}
\label{AB}
\end{figure}

Let's go on considering two subsystems $A$ and $B$ in the vacuum state. By conformal mapping we can always choose $B=[0,+\infty]$ and  $A=[-l_A-l_{A'},-l_{A'}]$. As shown in Fig.\ref{AB} we label $A'=[-l_{A'},0]$ and $B'=[-\infty,-l_A-l_{A'}]$. One can calculate the mutual information of $A,B$ and find the critical point $l_A=l_{A'}$.  \\
Let's focus on the case $l_A>l_{A'}$.. It is not hard to show in large $c$ limit $\rho_{A'B'}$ is a product state $\rho_{A'}\otimes \rho_{B'}$. This mean $A'$ and $B'$ lose correlations, so the quantum discord $\mathcal{Q}(B'|A')=\mathcal{Q}(A'|B')=0$ and classical correlation $\mathcal{C}(B'|A')=\mathcal{C}(A'|B')=0$. The state $\rho_{AA'B}=tr_{B'}|0\rangle\langle 0|$ is a mixed state, in general it can be written as the ensembles 
\bea
\rho_{AA'B}=\sum_i p_i |\psi\rangle_i ~_i\langle \psi|.
\eea
The decomposition is not unique. One can use the entanglement of formation (EoF) to characterize the quantum correlation between $A'$ and $AB$. For our case EoF is defined as \cite{Bennett:1996gf}
\bea\label{EoF}
E_f(A':AB)=\underset{p_i,|\psi\rangle_i}{\text{min}} \Big[\sum_i p_i S(\rho_{A',i}) \Big],
\eea
where $\rho_{A',i}:=tr_{AB}|\psi\rangle_i ~_i\langle \psi|$.
Unlike the quantum discord, we have $E_f(A':AB)=E_f(AB:A')$. \\
For any tripartite state $\rho_{A_1A_2A_3}$ such that $tr_{A_2}\rho_{A_1A_2A_3}=\rho_{A_1A_3}$ and $tr_{A_3}\rho_{A_1A_2A_3}=\rho_{A_1A_2}$,  we have the following monogamy relation, named Koashi-Winter relation\cite{Koashi},
\bea
E_f(A_1:A_2)+\mathcal{C}(A_1|A_3)\le S(\rho_{A_1}),
\eea
with equality if $\rho_{A_1A_2A_3}$ is pure. In the following we will use the Koashi-Winter relation for different subsystems.
Firstly, using the Koashi-Winter relation with $A_1=A'$, $A_2=B'$ and $A_3=AB$ we have the following equality,
\bea
E_f(A':B')+\mathcal{C}(A'|AB)= S(\rho_{A'}).
\eea
It is easy to show $E_f(A':B')=0$. By the definition of quantum discord we can obtain 
\bea
\mathcal{Q}(A'|AB)=I(A',AB)-\mathcal{C}(A'|AB)=S(\rho_{A'}).
 \eea
For a bipartite state, the quantum discord is equal to the entanglement entropy. So we have $\mathcal{Q}(A'|ABB')=S(\rho_{A'})=\mathcal{Q}(A'|AB)$. This is what we expect since $A'$ and $B'$ lose correlation, the measurement in $B'$ will not effect the reduced density matrix $\rho_{AA'B}$.   \\
Then, taking $A_1=A'$,$A_2=AB$ and $A_3=B'$ we have the following relation,
\bea\label{definitionEOF}
E_f(A':AB)+\mathcal{C}(A'|B')=S(\rho_{A'}).
\eea
Since $\mathcal{C}(A'|B')=0$ we have $E_f(A':AB)=S(\rho_{A'})$. Recall the definition of EoF (\ref{EoF}) the ensemble $\{p_i, |\psi\rangle_i\}$ that minimizes $\sum_i p_i S(\rho_{A',i})$  satisfies
\bea
\chi(\rho_{A'}):=S(\sum_i p_i\rho_{A',i})-\sum_i p_i S(\rho_{A',i})=0,
\eea
where $\rho_{A'}:=\sum_i p_i \rho_{A',i}$. That is the Holevo information of the mixed state $\rho_{A'}$ is vanishing, which means the states $\{\rho_{A',i}\}$ are  indistinguishable. Therefore, we have
\bea\label{purificationlike}
\rho_{A',i}:=tr_{AB}|\psi\rangle_i~_i\langle \psi|=\rho_{A'}.
\eea  
On the other hand we have $tr_{A'}\rho_{AA'B}=\rho_{AB}$, this leads to 
\bea
\sum_i p_i\rho^i_{AB}=\rho_{AB},
\eea
where $\rho^i_{AB}:= tr_{A'}|\psi\rangle_i~_i\langle\psi|$.  By the definition of EoF (\ref{definitionEOF})  we have
\bea
E_f(A':AB)=\sum_ip_iS(\rho^i_{AB})=S(\rho_{A'}).
\eea
From this one may calculate the Holevo informaiton of the ensemble $\rho_{AB}=\sum_i p_i\rho^i_{AB}$, 
\bea\label{HolevoAB}
\chi(\rho_{AB}):=S(\sum_i p_i\rho^i_{AB})-\sum_ip_iS(\rho^i_{AB})=S(\rho_{AB})-S(\rho_{A'})=S(\rho_{AA'B}).
\eea
 $\chi(\rho_{AB})$ is very large, which means at least some of the states $\{\rho^i_{AB}\}$ are distinguishable.  Moverover, for the ensemble $\rho_{AA'B}=\sum_i p_i |\psi\rangle_i ~_i\langle \psi|$, we have
\bea
\chi(\rho_{AA'B})=S(\rho_{AA'B}).
\eea
Therefore, we find
\bea\label{disconnectholevo}
\chi(\rho_{AB})=\chi(\rho_{AA'B}).
\eea
Coming back to the physical meaning of Holevo information, the above equality tells us the measurement in the region $A'$ cannot gain more information on the mixed state $\rho_{AA'B}$ in the ensemble $\sum_i p_i |\psi\rangle_i ~_i\langle \psi|$.\\
Finally, taking $A_1=A$, $A_2=A'B'$ and $A_3=B$, we have the equality,
\bea
E_f(A:A'B')+\mathcal{C}(A|B)=S(\rho_A).
\eea
Support the ensemble $\sum_i q_i |\phi\rangle_i ~_i\langle \phi|$ of the mixed state $\rho_{B'AA'}$ minimizes $\sum_i q_iS(\rho^i_{A'B'})$ or $\sum_i q_i S(\rho^i_{A})$, where $\rho^i_{A'B'}:= tr_{A}|\phi\rangle_i ~_i\langle \phi|$ and $\rho^i_A=tr_{A'B'}|\phi\rangle_i ~_i\langle \phi|$. We have
\bea\label{ensembleleft}
\rho_{A'B'}=\sum_i q_i \rho^i_{A'B'}=\rho_{A'}\otimes \rho_{B'},\quad \rho_{A}=\sum_i q_i \rho^i_A.
\eea
We can calculate the Holevo information of the above ensembles 
\bea\label{classicalcorrelationleft}
&&\chi(\rho_A):=S(\sum_i q_i \rho^i_A)-\sum_i q_i S(\rho^i_{A})=\mathcal{C}(A|B),\nonumber \\
&&\chi(\rho_{A'B'}):= S(\sum_i q_i \rho^i_{A'B'})-\sum_i q_i S(\rho^i_{A'B'})=S(\rho_{A'})+S(\rho_{B'})-S(\rho_A)+\mathcal{C}(A|B).
\eea
The Holevo information of the ensemble $\rho_{B'AA'}=\sum_i q_i |\phi\rangle_i ~_i\langle \phi|$ is given by the entanglement entropy $S(\rho_{B'AA'})$, i.e., $\chi(\rho_{B'AA'})=S(\rho_{B'AA'})= S(\rho_B)$.  With some calculations we obtain
\bea\label{quantumcorrelationleft}
\chi(\rho_{B'AA'})-\chi(\rho_{A'B'})=\mathcal{Q}(A|B).
\eea
If $A,B$ lose quantum correlation,  that is $\mathcal{Q}(B|A)=0$,  we may reproduce the result (\ref{disconnectholevo}). The quantum correlations between $A$ and $B$ can be understood as the difference between  the accessiable  information in $B'AA'$ and $A'B'$.
\subsubsection{Possible ansatz of the ensemble}
In last section we discuss some relations of the correlations among the subsystems in the vacuum state of holographic theory. These equalities do give us some physical explainations of the correlations between different subsystems.  However, they are useful to calculate the correlations only if we know the desired ensemble $\sum_i  p_i |\psi\rangle_i ~_i\langle \psi|$ and $\sum_i q_i |\phi\rangle_i ~_i\langle \phi|$.\\
From (\ref{purificationlike}) we can see the reduced density matrix by tracing over $AB$ of the pure states $|\psi\rangle_i$ are same. The pure states $|\psi\rangle_i$ can be seen as the purifications of the state $\rho_{A'}$.  Suppose $|\psi\rangle_0$ is one of the purification, one can construct other purfications by unitary operations working on $AB$, that is
\bea
|\psi\rangle_i= U^i_{AB}|\psi\rangle_0,
\eea
where $U^i_{AB}$ is the operator located in the region $AB$.
The ensemble is given by
\bea
\sum_i  p_i |\psi\rangle_i ~_i\langle \psi|=\sum_i p_i  U^i_{AB}|\psi\rangle_0 ~_0\langle\psi|(U^i_{AB})^\dagger.
\eea
The ansatz should satisfy some constraints such as $S(\rho_{AA'B})+S(\rho_{A'})=S(\rho_{AB})$, but at present we cannot find a way to  fix the coefficients $p_i$ and unitarty operations $U^i_{AB}$.  Perhaps, it can be associated with the idea of state/surface correspodence\cite{Miyaji:2015yva}. We leave this to the future works. \\

\section{Conclusion}\label{conclusion}
In this paper we discuss correlations in the geometric states.  Using the upper bound of Holevo information we find the set of geometric states cannot be convex. Convex combination gives us a way to construct non-geometric state.  In \cite{Guo:2018fnv} we show the superposition of two pure geometric states cannot be geometric. In that case the criterion is that the subsystem $A$ in the superposition state has entanglement entropy of order $c^2$, which is not possible for a holographic theory. But here the entanglement entropy of $A$ in the state $\rho_c^g$ is still of order $c$.  The underlying reason may be the non-linearity of general relativity. In this paper we assume the Shannon entropy $H(p_i)$ is of order $c^0$.  If the number of the combination is of order $c$ such that $H(p_i)\sim O(c)$, the result should be modified. We will explore on this topic in the near future.\\ 
We construct several states with corrrelation between two CFTs. All these examples $\rho_{I,\beta}$, $|\Psi\rangle_\lambda$ and $\rho_{II,\beta}$ fail to be dual to a connected spacetime because the correlation between two CFTs is almost vanishing. The state $|\Psi\rangle_\lambda$ is interesting, which is  very similar as the thermofield doulbe state. The microcanonical ensemble is indistinguishable for measurement located a small region comparing with the whole system. But their difference will appear if the measurements are performed in a larger subregion. One may use the R\'enyi entropy as a probe to detect their difference\cite{Dong:2018lsk}\cite{Guo:2018fye}.  In \cite{Marolf:2018ldl} the author construct a modified state of $|\Psi\rangle_\lambda$ 
\bea
|\psi\rangle=\sum_{E_i}e^{-\beta E_i/2}f(E_i-E)|E_i\rangle_1|E_i\rangle_2,
\eea
where $f(x)$ is a function that is sharply peaked at $f(0)=1$. It is expected the two point correlator $\langle \psi| O_1 O_2 |\psi\rangle$ is nonvanishing if energy-width is non-zero. It is an interesing question to evaluate the two point correlator and compare with the proposed geoemtry dual to the state $|\psi\rangle$.\\ 
To quantify the relation between correlations and geometry it is necessary to divide the correlations into classical and quantum ones.  In this paper we try to use the geometric measures of quantum discord and classical correlation. But we only can perform the calculations for some pure states. The non-trivial examples of mixed  state will give us more insight on important role of  classical and quauntum correlation   in a given geometric state. This is a question that is worth to explore more in the future. \\
We also consider two intervals $A$ and $B$ with distance $d$ in the vacuum state of 2D CFTs.  The feature of CFTs with large $c$ is  the phase transition of mutual information of $I(\rho_{AB})$ with respect to distance $d$. If the distance is larger than the critical point $l_c$, the reduced density matrix $\rho_{AB}$ is quasi-product, that is $\rho_{AB}=\rho_A\otimes\rho_B+\delta\rho_{AB}$. We find the main contribution of the $I(\rho_{AB})$ is near the spectrum $\lambda_0=e^{-2(b_{A}+b_B)}$ where $b_{A(B)}=\frac{c}{6}\log l_{A(B)}/\epsilon$. The perturbation of spectrum $\bar \delta_0$ from $\delta\rho_{AB}$ should be suppressed by $\frac{1}{c}$.  
But we haven't sucessfully expain why there is a critical point of $d$ because of the lack of enough information on the dependence of the perturbation $\delta\rho_{AB}$ on distance $d$. \\
We also analyse the correlation relations of the subsystems $A$,$B$ and the complementary $A'$,$B'$ by using the Koashi-Winter relation of any tripartite states. The classical and quantum correlation between $A$ and $B$ can be expressed as Holevo information. This helps us to understand more on the physical meaning of the correlations between $A$ and $B$.  A more important question is how to sew the states of $\rho_{A}$, $\rho_{A'}$ and $\rho_{B}$, $\rho_{B'}$ together such that they show the correlations as the geometric states. Of course we are still far away from solving the problem.  But we hope the results in this paper may help us to catch a glimpse on the correlations in geometric states.

 \section*{Acknowledgement}
I am supported by the Fundamental Research Funds for the Central Universities under Grants NO.2020kfyXJJS041.
\appendix
\section{Holevo information in 2D CFTs}\label{appendixholevo}
In this appendix we will briefly review how to calculate Holevo information in 2D CFTs. Interesting readers may refer to 
\cite{Guo:2018djz} for more details. We only consider the contributions from the vacuum conformal family, the entanglement entropy of a short interval $[0,\ell]$ in a translation invariant state $\rho$  up to $O(\ell^{12})$ is\cite{Guo:2018pvi}
\bea \label{EEgeneral}
&& S_A = \frac{c}{6}\log\frac{\ell}{\epsilon} + \ell^2 a_T \Trho
                                  + \ell^4 a_{TT} \Trho^2
                                  + \ell^6 a_{TTT} \Trho^3 \nonumber \\
&& \phantom{S_A =}
           + \ell^8 \big( a_{\mA\mA} \Arho^2
                        + a_{TT\mA} \Trho^2\Arho
                        + a_{TTTT} \Trho^4 \big) \nonumber\\
&& \phantom{S_A =}
           + \ell^{10} \big( a_{T\mA\mA} \Trho\Arho^2
                           + a_{TTT\mA} \Trho^3\Arho
                           + a_{TTTTT} \Trho^5 \big) \nonumber\\
&& \phantom{S_A =}
           + \ell^{12} \big( a_{\mB\mB} \Brho^2
                           + a_{\mD\mD} \Drho^2
                           + a_{T\mA\mB} \Trho\Arho\Brho
                           + a_{T\mA\mD} \Trho\Arho\Drho \nonumber\\
&& \phantom{S_A =}
                           + a_{\mA\mA\mA} \Arho^3
                           + a_{TTT\mB} \Trho^3\Brho
                           + a_{TTT\mD} \Trho^3\Drho
                           + a_{TT\mA\mA} \Trho^2\Arho^2 \nonumber\\
&& \phantom{S_A =}
                           + a_{TTTT\mA} \Trho^4\Arho
                           + a_{TTTTTT} \Trho^6  \big) + O(\ell^{14}),
\eea
where $T$,$\mA$,$\mathcal{B}$ and $\mathcal{D}$ are the quasiprimary operators, $a_T$,$a_{TT}$...are constant coefficients\cite{Guo:2018pvi}. From (\ref{holevodefinition}) to calculate Holevo information we need to evaluate the averge entanglement entropy $\sum_i p_i S(\rho_i)$, which is associated with the average one-point functions such as
\bea
\sum_i p_i \langle T\rangle_i^2,\quad \sum_i p_i \langle T\rangle_i \langle \mA \rangle_i,
\eea
and so on. The Holevo information of a short interval in the canonical ensemble state with inverse temperature $\beta$ is an example that is shown in in \cite{Guo:2018djz}.
For $\mathcal{X}=T,\mA$ and $\mathcal{Y}=T,\mA$, it can be shown
\bea
p_i \langle \mathcal{X}\rangle_i \langle \mathcal{Y}\rangle_i =\frac{1}{L}\int_{-\frac{L}{2}}^{\frac{L}{2}} dx \langle \mathcal{X}(x)\mathcal{Y}\rangle_\beta,
\eea
where $p_i=e^{-\beta E_i}/Z(\beta)$. Using this formula one can evaluate the Holevo information up to order $O(\ell^{10})$. But for higher order quasiprimary operators $\mB$ and $\mD$, we cannot use this formula \cite{Guo:2018djz}. To calculate the order $O(\ell^{12})$ we need
\bea\label{appendix1}
\sum_i p_i \langle \mB\rangle_i^2,\quad \sum_i p_i \langle \mD \rangle_i^2,\quad \sum_i p_i \langle T\rangle_i \langle \mA\rangle_i \langle \mB \rangle_i,...,
\eea
 and so on. For the microcanonical ensemble $p_i= \frac{\delta(E_i-E)}{\Omega(E)}$, let's firstly consider 
\bea
\mathcal{G}_{\mathcal{X}_1\mathcal{X}_2...\mathcal{X}_k}(\lambda):=\frac{1}{\Omega(E)}\sum_i \langle E_{i_1}| \mathcal{X}_1 |E_{i_2}\rangle \langle E_{i_2} |\mathcal{X}_2|E_{i_3}\rangle...\langle E_{i_{k}} |\mathcal{X}_k|E_{i_1}\rangle\delta(E_{i_1}-E)\delta(E_{i_2}-E)...\delta(E_{i_k}-E),\nonumber
\eea
with $\mathcal{X}_1,\mathcal{X}_2= T, \mA, \mB, \mD$. If $\mathcal{X}_1,\mathcal{X}_2= T,\mA$, or one of the $\mathcal{X}_k$ is $\mB$ or $\mD$, say $\mathcal{X}_1$, the expression can be associated with the average one-point functions (\ref{appendix1}). $|E_{i_j}\rangle$ are in same energy level, so only zero mode of $T$ and $\mA$ give the contributions to expectation value. Zero mode of $T$ and $\mA$ are commutative, the states $E_{i_j}$ can be organized as the common eigenstates of $T$ and $\mA$. Therefore,  with $\mathcal{X}_i=T, \mA$ or one of the $\mathcal{X}_k$ is $\mB$ or $\mD$, $\mathcal{G}_{\mathcal{X}_1\mathcal{X}_2...\mathcal{X}_k}(\lambda)$ is equal to 
\bea
\frac{1}{\Omega(E)}\sum_{i}\langle \mathcal{X}_1\rangle_i \langle \mathcal{X}_2\rangle_i...\langle \mathcal{X}_k\rangle_i\delta(E_i-E),
\eea
where $\langle \mathcal{X}_j\rangle_i:=\langle E_i|\mathcal{X}_j |E_i\rangle$. We are interested in the thermodynamic limit $L\to\infty$. It can be shown 
\bea\label{appendix2}
\lim_{L\to \infty} \mathcal{G}_{\mathcal{X}_1\mathcal{X}_2...\mathcal{X}_k}(\lambda)= \langle \mathcal{X}_1\rangle_\lambda\langle \mathcal{X}_2\rangle_\lambda...\langle \mathcal{X}_k\rangle_\lambda,
\eea
where $\langle \mathcal{X}_j\rangle_\lambda:=\frac{1}{\Omega(E)}\sum_i \langle \mathcal{X}_j\rangle_i$. By using the definition of (\ref{holevodefinition}), the calculation of Holevo information will involve the terms
\bea
a_{\mathcal{X}_1\mathcal{X}_2...\mathcal{X}_k}\Big(\langle \mathcal{X}_1\rangle_\lambda...\langle \mathcal{X}_k\rangle_\lambda-\frac{1}{\Omega(E)}\sum_i \langle \mathcal{X}_1 \rangle_i...\langle \mathcal{X}_k \rangle_i\delta(E_i-E)\Big). 
\eea
By using (\ref{appendix2}) we find all of the terms are vanishing except 
\bea
a_{\mB\mB}\Big( \langle \mB\rangle_\lambda^2 -\frac{1}{\Omega(E)}\sum_i \langle \mB\rangle_i^2\Big)+a_{\mD\mD}\Big( \langle \mD\rangle_\lambda^2 -\frac{1}{\Omega(E)}\sum_i \langle \mD\rangle_i^2\Big).
\eea
One could also check the Holevo information is vanishing up to $O(\ell^{10})$ by directly taking the limit $L\to \infty$ of the exact results in \cite{Guo:2018djz}.

\end{document}